\documentclass[aps,prb,twocolumn,amsmath,amssymb,superscriptaddress,floatfix]{revtex4-2}
\pdfoutput=1

\usepackage{mathrsfs}
\usepackage{amsmath}
\usepackage{mathtools}
\usepackage{braket}
\usepackage{amsfonts}
\usepackage{float}
\usepackage{graphicx} 
\usepackage[caption=false]{subfig}
\usepackage{bm} 
\usepackage[dvipsnames]{xcolor}
\usepackage{epstopdf}
\usepackage{comment}
\usepackage{soul}
\usepackage{multirow}

\newcommand\red[1]{\textcolor{black}{#1}}
\newcommand\blue[1]{\textcolor{black}{#1}}
\newcommand\green[1]{\textcolor{black}{#1}}

\begin{document}
\title{Entanglement in quenched extended Su-Schrieffer-Heeger model with anomalous dynamical quantum phase transitions}

\author{ Cheuk Yiu Wong }
\thanks{These authors contributed equally to this work.}
\affiliation{ \textit Department of Physics, City University of Hong Kong, Kowloon, Hong Kong, China}

\author{ Tsz Hin Hui }
\thanks{These authors contributed equally to this work.}
\affiliation{ \textit Department of Physics, City University of Hong Kong, Kowloon, Hong Kong, China}

\author{P. D. Sacramento}
\affiliation{ \textit CeFEMA, Instituto Superior T\'ecnico, Universidade de Lisboa, Av. Rovisco Pais, 1049-001 Lisboa, Portugal}

\author{ Wing Chi Yu }
\email{wingcyu@cityu.edu.hk}
\affiliation{ \textit Department of Physics, City University of Hong Kong, Kowloon, Hong Kong, China}

\date{\today}

\begin{abstract}

Research on topological models unveils fascinating physics, especially in the realm of dynamical quantum phase transitions (DQPTs). However, the understanding of entanglement structures and properties near DQPT in models with longer-range hoppings is far from complete. In this work, we study DQPTs in the quenched extended Su-Schrieffer-Heeger (SSH) model. Anomalous DQPTs, where the number of critical momenta exceeds the winding number differences between the pre-quench and post-quench phases, are observed. We find that the entanglement exhibits local maximum (minimum) around the anomalous DQPTs, in line with the level crossings (separations) around the middle of the correlation matrix spectrum. We further categorize the phases in the equilibrium model into two classes and distinctive features in the time evolution of the entanglement involving quenches within and across the two classes are identified. The findings pave the way to a better understanding of topological models with longer-range hoppings in the out-of-equilibrium regime.

\end{abstract}

\maketitle

\section{Introduction}


\red{The formal establishment of the theory of dynamical quantum phase transitions (DQPTs) in 2013 \cite{Heyl2013}, followed by the studies of its fundamental properties \cite{Heyl2015,Heyl2014,Weidinger2017,Corps2022,Corps2023} and successful experimental realization \cite{Peng2015,Jurcevic2017,Tian2019,Flaschner2018}, leads to an alternative approach in studying non-equilibrium physics in quantum many-body systems. DQPTs have been witnessed and substantially studied in spin models \cite{Titum2019,Jafari2019,Schmitt2018,Vosk2014,Nicola2020,Nicola2021,HeylPollmann2018,Vajna2014,Uhrich2020,Andraschko2014,Lacki2019,Zvyagin2017,Bandyopadhyay2021,Markov2021,Ding2020,Karrasch2013,Kriel2014,Cheraghi2020,Kennes2018,Seetharam2021,Lang2018,Porta2020,Wong2022,Wong2023,Cheraghi2023,Khan2023_1,Khan2023_2,typeBanomalous,Stumper2022,sharma2015,Niu2023,Haldar2020}, topological models \cite{Schmitt2015,Budich2016,Sedlmayr2018,JafariJohannesson2019,Mishra2020,Yu2021,Poyhonen2021,Sadrzadeh2021,Zeng2023,Zhang2022,Hou2022,accident2015,multibandSSHDQPT,NNNSSHDQPT,Huang2020,rossi2022,Lakkaraju2024} and high energy physics \cite{Zache2019,Mueller2023,Lahiri2019,Damme2022,Pedersen2021,Osborne2024,Knaute2023}, via various protocols such as} \green{a quench or a} \red{periodical drive \cite{YangZhou2019,Kosior2018,Kyaw2020,Jafari2021,Zamani2020,Jafari2022,Naji2022,Naji2023} to bring the system out of equilibrium.} There are three types of DQPTs. The type-I  DQPTs (DPTs-I) considers 
the kinks at the transition point in the long-time average of the order parameter of a system \red{\cite{Corps2022,Corps2023,Zunkovic2018,Halimeh2017,Zunkovic2016}}. On the other hand, the type-II DQPTs (DPTs-II), \green{ which will be the main focus of this work,} concerns the transient time dynamics, and are defined by the zeros of the Loschmidt amplitude (LA), the overlap of the time-evolved state driven by the unitary evolution of the final Hamiltonian $H_f$ onto the initial state $\ket{\psi_0^i}$
\begin{equation} \label{eqn:LA}
    \mathcal{G}_0(t) = \braket{\psi_0^i | e^{-iH_ft} | \psi_0^i},
\end{equation}
and in turn the nonanalyticities in the Loschmidt rate (also known as the dynamical free energy) $\lambda_0(t) = -\lim_{N \rightarrow \infty} N^{-1}\ln|\mathcal{G}_0(t)|^2$ with $N$ the system size \cite{Heyl2013,Heyl2018,Zvyagin2016}. Recently, a third type of DQPT (DPTs-III) has also been introduced as the change in the scaling behavior of a physical observable upon approaching the steady state in the long-time limit \cite{Ding2023,Aditya2022,Nandy2018,Sen2016}. 

\begin{table*}[]
\begin{tabular}{|c|cc|cc|}
\hline
                            & \multicolumn{1}{c|}{$W_1 \leftrightarrow W_0$ regular}          & $W_2 \leftrightarrow W_2$ regular        & \multicolumn{1}{c|}{$W_1 \leftrightarrow W_0$ anomalous}                   & $W_2 \leftrightarrow W_2$ anomalous              \\ 
\hline
    $k^*$ & \multicolumn{1}{c|}{$1k^*$}                   & $0k^*$                 & \multicolumn{1}{c|}{$3k^*$, extra $k^*$ from conjugate pair} & $2k^*$, extra $k^*$ from conjugate pair \\ 
\hline
$t^*$    & \multicolumn{1}{c|}{relatively longer}                                     & \multicolumn{1}{c|}{N.A.}& \multicolumn{2}{c|}{relatively shorter}\\  
\hline
    $\vec{r}_k$                       & \multicolumn{1}{c|}{$1k^*$ in 1 loop}         & N.A.                & \multicolumn{2}{c|}{$k^*$ pair in 1 loop}                                                     \\ 
\hline
    $S_A$        & \multicolumn{1}{c|}{sometimes extrema around $t^*$}                      &     N.A.                & \multicolumn{1}{c|}{first extrema around $t^*$}    &      near extrema around $t^*$                                                    \\ 
\hline
    $\{\xi_j\}$ & \multicolumn{2}{c|}{eigen-levels around Fermi level are \red{gapped}} & \multicolumn{2}{c|}{eigen-levels tend to mix}                                                 \\ 
\hline
\end{tabular}
	\caption{Summary of the features of the critical momentum $k^*$, critical time $t^*$, the dynamical vector $\vec{r}_k$, entanglement entropy $S_A$, correlation matrix spectrum $\{\xi_j\}$ evolution in different dynamical phases. The regular and anomalous phase refer to the case where the number of critical momenta is equal to and larger than the absolute winding number difference, respectively (see the text for the details). ``N.A." in the table stands for ``not applicable" since there is no DQPT.} 
	\label{tab:sum}
\end{table*}

\green{The type-II DQPTs are commonly found in models being quenched across the underlying equilibrium phase boundary, however, exceptions exist \cite{Jafari2019,Vajna2014,Andraschko2014,Stumper2022,sharma2015,Sadrzadeh2021}. 
On the other hand, DQPTs can also occur for quench within the same phase \cite{Jafari2019,Uhrich2020,Uhrich2020,Vajna2014,Andraschko2014,typeBanomalous,sharma2015,accident2015,NNNSSHDQPT,rossi2022}. The connection, if any, between the equilibrium and the dynamical phase diagrams remain to be addressed.}

In topological models, it is found that there can be more than one critical momenta \green{(and correspondingly critical time scales)} \cite{Budich2016,accident2015,multibandSSHDQPT,NNNSSHDQPT,Huang2020,rossi2022}. \red{These critical momenta originate from the different topological structures between the initial and final phases, in which their number must be at least the difference in the winding number between the two phases for 1D models. The cases with excessive critical momenta are referred to as accidental cases in Ref. \cite{accident2015}. For 2D topological models, robust DQPTs are present if there are at least one node in the $k$-space wavefunction overlap, and such nodes are guaranteed when the pre- and postquenched phases have a different Chern number \cite{NNNSSHDQPT,Huang2020}.} \green{The 2D case in general has areas of zeros of the Loschmidt amplitude distributed on the complex time plane and a continuum of solutions of the critical times, but it is found in a number of cases that have line-like distribution of zeros whose critical time usually agrees well with the Loschmidt rate non-analytic peak \cite{pedro2024}.} 

\red{In this work, we show that the previously referred to accidental DQPTs in topological models} \green{in fact contains three notions, distinguished by quenches (1) within the same phase, (2) across the phase boundary but without a change in the winding numbers, and (3) across the boundary with a change in the winding numbers.} We demonstrate our findings via an extended version of the Su-Schrieffer-Heeger (SSH) model, a commonly studied topological model in condensed matter physics, due to its simplicity and accessibility to experiments \cite{experimentSSH}. \green{The extended SSH model possesses a rich equilibrium phase diagram, allowing us to probe the quantum dynamics of quenches involving different topological and the trivial phases.}
\red{Specifically, we realize for quenches between trivial and the topological phases, and between phases with the same winding number, there exists regions in the dynamical phase diagram, where the number of critical momenta $n_{k^*} > |\Delta W|$, the absolute difference of winding number between the two phases. Distinctive} features are found between the cases of $n_{k^*} = |\Delta W|$ and $n_{k^*} > |\Delta W|$. \green{For quench between phases with the same winding number, DQPTs are not just found to be possible for quench within the same phase, but also across the equilibrium phase boundary.} \red{We reckon the observations may be a generic nonequilibrium feature for topological models beyond nearest-neighbor hoppings, and further formulate and term these kinds of accidental DQPTs the \textit{type-B anomalous} DQPTs.} Note that the type-B anomalous DQPTs here is different from the anomalous DQPTs (we call them type-A anomalous DQPTs for unambiguity) described in \blue{Ref. \cite{Corps2022,Halimeh2017,Homrighausen2017,Zauner-Stauber2017,Halimeh2021,Corps2023,Nicola2021,Halimeh2020,Hashizume2022,Osborne2024}}, which emphasizes the minimum in rate functions before the first critical time. 

\green{We further investigate the origin of the critical momenta and the time evolution of the entanglement entropy (EE) and the correlation matrix spectrum in type-B anomalous DQPTs. The main findings are summarized in Table \ref{tab:sum}.} It is found that the extra $k^*$'s in the anomalous regions are in fact complex conjugate pair of the roots from the orthogonality of the initial and final Bloch vectors. Motivated by the distinctive behavior of the EE in the equilibrium case, we categorize the four phases in the extended SSH model into two classes according to the extent of non-locality of the ground state \green{to extract the generic features in the EE.}  We considered \red{a variety of} quenches within a class and across two classes, \red{and the EE extremum properties near DQPTs in accordance with the regular ($n_{k^*} = |\Delta W|$) and type-B anomalous ($n_{k^*} > |\Delta W|$) regions are presented in Table \ref{tab:ent}.} 
Furthermore, distinguishing features in the correlation matrix spectrum are also found between the regular and type-B anomalous dynamical phases. In the latter case, a significantly larger number of eigen-levels are involved in the evolution in the long-time regime. 


This paper is organized as follows: Sec. \ref{sec:ks_diag} demonstrates the dynamical phase diagram of some selected initial points in each equilibrium phase of extended SSH model, for which we provide analytical characteristics and reasoning as to how the \red{type-B }anomalous regions emerge and their distinction to the regular case. We further discuss the differences between type-B anomalous and regular regions regarding the entanglement entropy in the vicinity of DQPTs in different quench scenarios in Sec. \ref{sec:EE}, and the time evolution of the correlation matrix spectrum in Sec. \ref{sec:CMS}. Finally, we give a conclusion and outlook for future study in Sec. \ref{sec:conclusion}. 

			
			
			
			






  
			

\section{Dynamical Phase Diagram of the extended SSH model}
\label{sec:ks_diag}

\begin{figure}
    \centering
    \includegraphics[width=8cm]{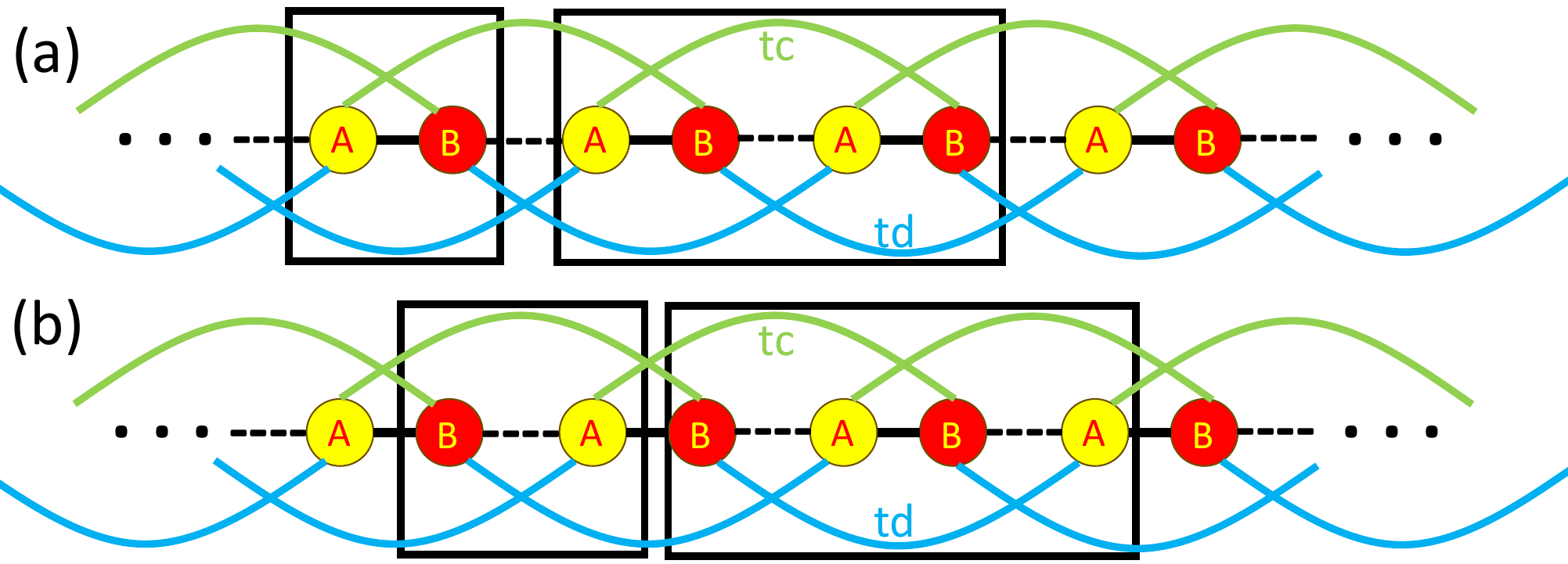}
    \caption{Schematic diagram of the extended SSH chain, with different subsystems \red{in (a) $L_A = 1$ (small box) and $L_A = 2$ (large box), (b) $L_B = 1$ (small box) and $L_B = 2$ (large box). $L_{A/B}$ denotes the number of unit cells} 
    starting at A/B site from the left side. Green and blue curves denote the $t_c$ and $t_d$ hopping terms respectively, and the black solid and dashed lines denote $t_a$ and $t_b$ respectively. }
    \label{fig:chain}
\end{figure}

The Hamiltonian of the extended SSH model reads \red{\cite{Maffei2018}}
\begin{align}
    H &= \sum_{j = 1}^L ( t_ac_{j,A}^\dagger c_{j,B} + t_bc_{j,B}^\dagger c_{j + 1,A} \nonumber \\
    & \qquad + t_cc_{j,A}^\dagger c_{j + 1,B} + t_dc_{j,B}^\dagger c_{j + 2,A} + \text{h.c.} ),
\label{eqn:H}
\end{align}
where \textit{L} is the number of unit cells, \textit{A} and \textit{B} denote the two sublattices, $t_a = -t( 1 + \eta )$ and $t_b = -t( 1 - \eta )$ are the intra- and inter-cell hoppings in the original SSH model. The $t_c$ and $t_d$ terms are added to include chiral-symmetry-preserving next-nearest-neigbhor hoppings (see Fig. \ref{fig:chain} for an illustration of the model). 
The model possesses topological phases with winding numbers $W=1,-1$, and 2 respectively, and a trivial phase with winding number $W=0$, in which we denote them as $W_0$, $W_1$, $W_2$ and $W_{-1}$ in the remaining text. Without the loss of generality, we will consider $\eta=0, t=1$, and adopt periodic boundary condition in the following, unless otherwise specified. 

The model can be solved exactly via Fourier transformation as
\begin{align}
    a_k &= \frac{1}{\sqrt{L}} \sum_j e^{-ikj} c_{j,A} \\
    b_k &= \frac{1}{\sqrt{L}} \sum_j e^{-ikj} c_{j,B}
\end{align}
which brings the Hamiltonian of the system into the form of 
\begin{equation}
    H = \sum_k \Psi_k^\dagger H(k) \Psi_k,
\end{equation}
where $\Psi_k = ( a_k,b_k )^T$ and $H(k)$ is the Bloch Hamiltonian $H(k) = \vec{d}(k) \cdot \vec{\sigma}$ where the associated Bloch vector has the components
\begin{equation}
    \begin{aligned}
        d_x(k) &= t_a + ( t_b + t_c )\cos k + t_d\cos( 2k ) \\
        d_y(k) &= ( t_b - t_c )\sin k + t_d\sin( 2k ) \\
        d_z(k) &= 0
    \end{aligned}
\end{equation}
and $\vec{\sigma}$ is the Pauli matrix vector.

The system is first prepared in the ground state $\ket{\psi_0^i}$ of the Hamiltonian $H_i=H(t_c^i,t_d^i)$, then is brought out of equilibrium by a sudden quench and allowed to evolve with the Hamiltonian $H_f=H(t_c^f,t_d^f)$. Throughout the paper, we use the simplified notation $(t_c^i,t_d^i) \rightarrow (t_c^f,t_d^f)$ to indicate a quench. The Loschmidt amplitude in two-band models can be analytically calculated as \cite{Budich2016}
\begin{equation}
    \mathcal{G}_0(t) = \prod_k \mathcal{G}_k(t) = \prod_k ( |g_k|^2  e^{i\varepsilon_k^f t} + |e_k|^2 e^{-i\varepsilon_k^f t} )
\end{equation}
with $|g_k|^2 = ( 1 + \hat{d}_i(k) \cdot \hat{d}_f(k) ) / 2$ and $|e_k|^2 = ( 1 - \hat{d}_i(k) \cdot \hat{d}_f(k) ) / 2$, $\hat{d}(k) = \vec{d}(k)/|\vec{d}(k)|$ is the unit Bloch vector and the quasiparticle energy $\varepsilon_k^f = \pm|\vec{d}_f(k)|$. The orthogonality between the two Bloch vectors, i.e. $\vec{d}_i(k^*) \cdot \vec{d}_f(k^*)=0$ results in a cubic equation of $\cos k$ that defines the critical momenta $k^*$ associated with the critical times 
\begin{eqnarray}
t_n^*=\frac{\pi}{\varepsilon_{k^*}^f}\left(n-\frac{1}{2}\right), 
\label{eq:tstar}
\end{eqnarray}
where $n=1,2,\cdots$, when DQPTs take place. The number of real critical momenta $n_{k^*}$ ranges from 0 to 3, depending on the quench parameters. In the following discussion, the critical times are obtained from Eq. (\ref{eq:tstar}) in the thermodynamic limit.

Alternatively, one also detects DQPTs by the dynamical topological order parameter (DTOP) defined as \cite{Budich2016}
\begin{equation}
    \nu_D(t) = \frac{1}{2\pi}\int_{0}^{\pi} dk\frac{\partial \phi_k^G(t)}{\partial k},
\end{equation}
where $\phi_k^G(t) = \phi_k^{\mathcal{G}_0}(t) - \phi_k^\text{dyn}(t)$ is the Pancharatnam geometric phase (PGP) \cite{Pancharatnam1956,Samuel1988}. $\phi_k^{\mathcal{G}_0}(t)$ is the argument of $\mathcal{G}_k(t)$ and the dynamical phase takes the form
\begin{equation}
    \begin{aligned}
        \phi_k^\text{dyn}(t) &= -\int_0^t dt' \braket{\psi_k^i(t') | H_f(k) | \psi_k^i(t')} \\
        &= ( |g_k|^2 - |e_k|^2 ) \varepsilon_k^f t.
    \end{aligned}
\end{equation}
DTOP usually jumps by an integer at critical times while stays constant for $t \neq t_n^*$. In some quench scenarios in some models, the jumps can be of half-integers \cite{Ding2020,Wong2023}. 
The jump in DTOP can be positive or negative depending on the critical momenta. 
The dynamics of the  $\mathcal{G}_k(t)$, combined with the PGP, can be visualized by the dynamical vector \cite{Ding2020}
\begin{equation}
    \vec{r}_k(t) = ( x_k(t),y_k(t) ) \leftrightarrow |\mathcal{G}_k(t)| e^{i\phi_k^G(t)}.
\end{equation}
It was shown that the trajectory of $\vec{r}_k$ in the first Brillouin zone at the first critical time reveals distinct features for quenches from the equilibrium phase boundary \cite{Ding2020}. Here, we find the trajectories of $\vec{r}_k$ exhibit different characteristics for the quenches involving the regular and anomalous regions in the dynamical phase diagram, and the results will be presented below.

\begin{figure}
    \centering
\includegraphics[width=8cm]{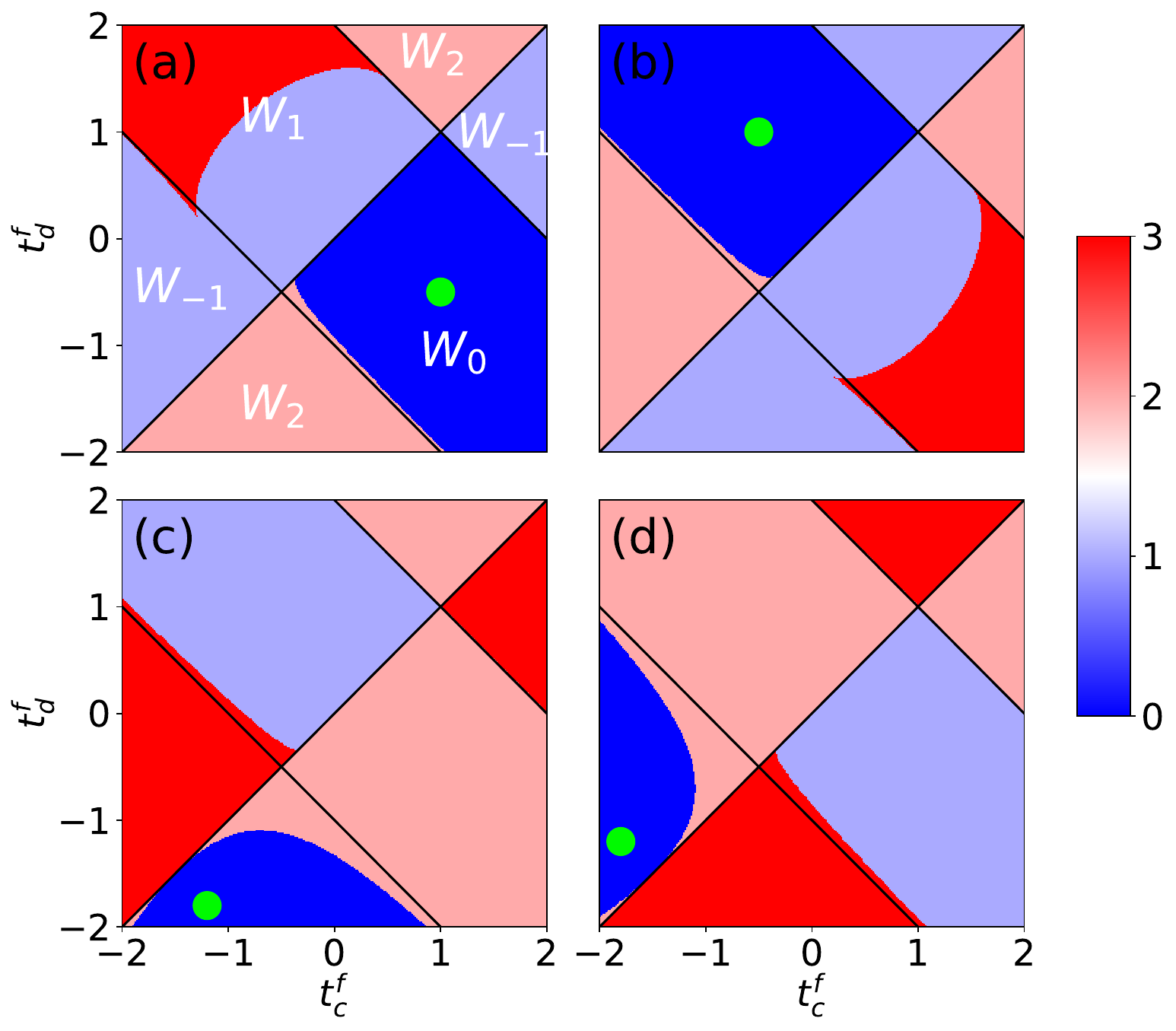} 
    \caption{Dynamical phase diagrams characterized by the number of critical momenta $n_{k^*}$ indicated by the color bar in the extended SSH model with the initial state (the green dot) prepared in the (a) $W_0$, (b) $W_1$, (c) $W_2$ and (d) $W_{-1}$ phase. The black solid lines show the equilibrium phase boundaries.}
    \label{fig:Dphasediagram}
\end{figure}

\begin{figure}
    \centering
    \includegraphics[width=8.5cm]{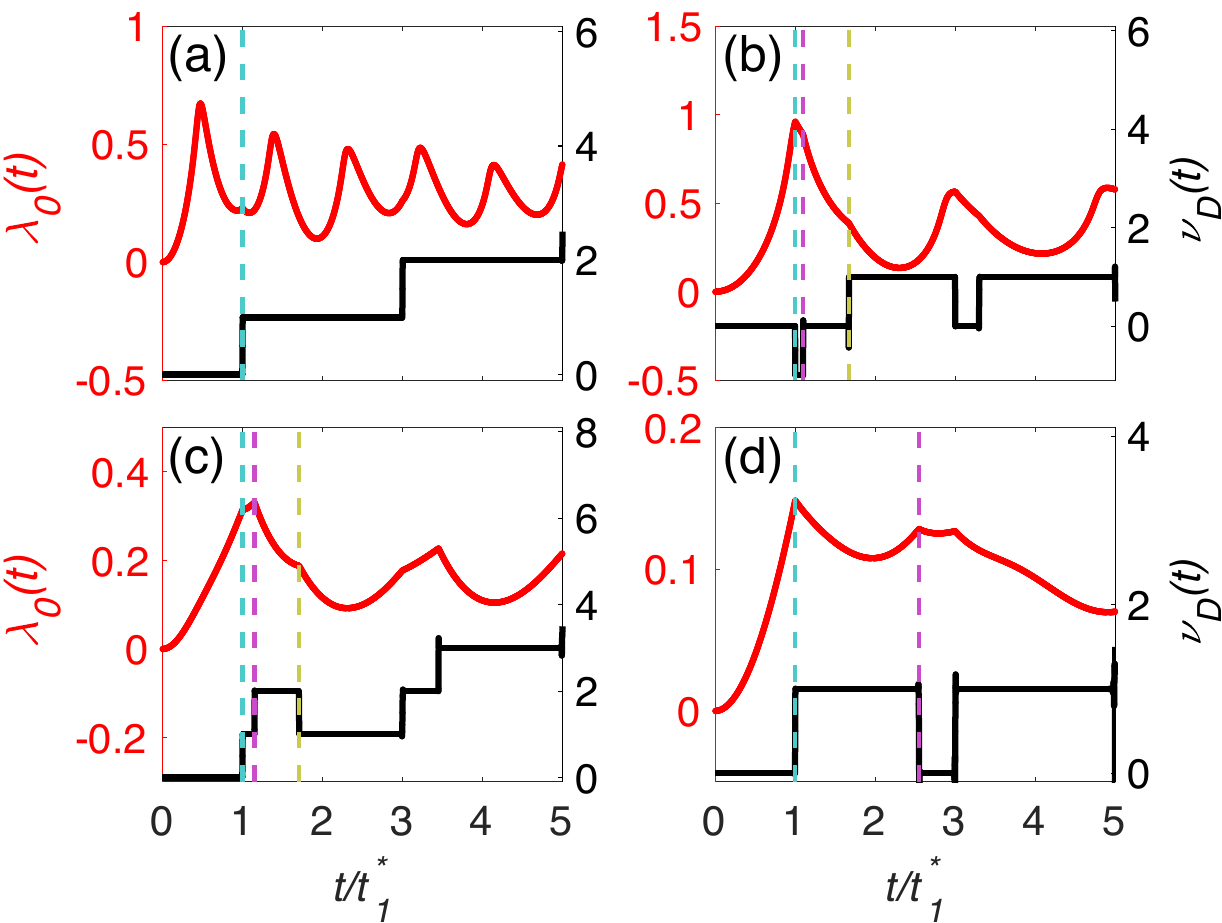}
    \caption{Demonstrations of the Loschmidt rates (red lines) and DTOPs (black lines) of the quenches (a) $(-0.5,1) \rightarrow (0.8,-0.8)$ ($W_1 \rightarrow W_0$ 1$k^*$ region), (b) $(-0.5,1) \rightarrow (1.5,-1)$ ($W_1 \rightarrow W_0$ 3$k^*$ region), (c) $(-1.2,-1.8) \rightarrow (-1.7,-0.6)$ (bottom $W_2 \rightarrow$ left $W_{-1}$), and (d) $(-1.2,-1.8) \rightarrow (-0.4,-1)$ (bottom $W_2 \rightarrow$ bottom $W_2$ with DQPT), in an $L = 8000$ system. Colored dashed lines indicate first appearance of critical times.}
    \label{fig3_LR_1ks3ks}
\end{figure}

\begin{figure}
    \centering
    \includegraphics[width=9cm]{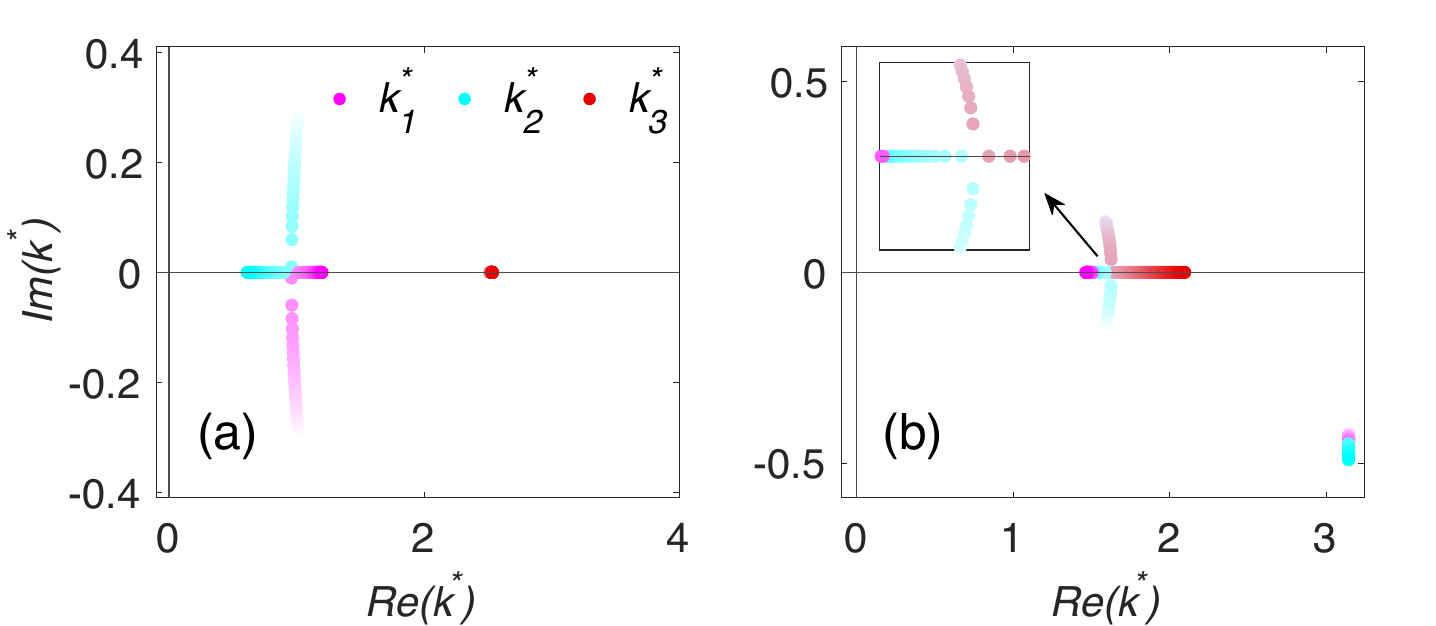}
    \caption{Locations of $k^*$'s in the complex plane in the vicinity of the anomalous dynamical phase boundaries of an $L = 400$ system for quenches (a) $t_c^i = -0.5,t_d^i = 1$ to $t_c^f$ ranging from 1 to 1.6 with $t_d^f = -0.7$ and (b) $t_c^i = -1.2,t_d^i = -1.8$ to $t_d^f$ ranging from $-1.4$ to $-0.8$ with $t_c^f = -0.2$. Color dots from light to dark colors represent increasing final parameters.}
    \label{fig4_ks_traj}
\end{figure}

\begin{figure}
    \centering
    \includegraphics[width=8.5cm]{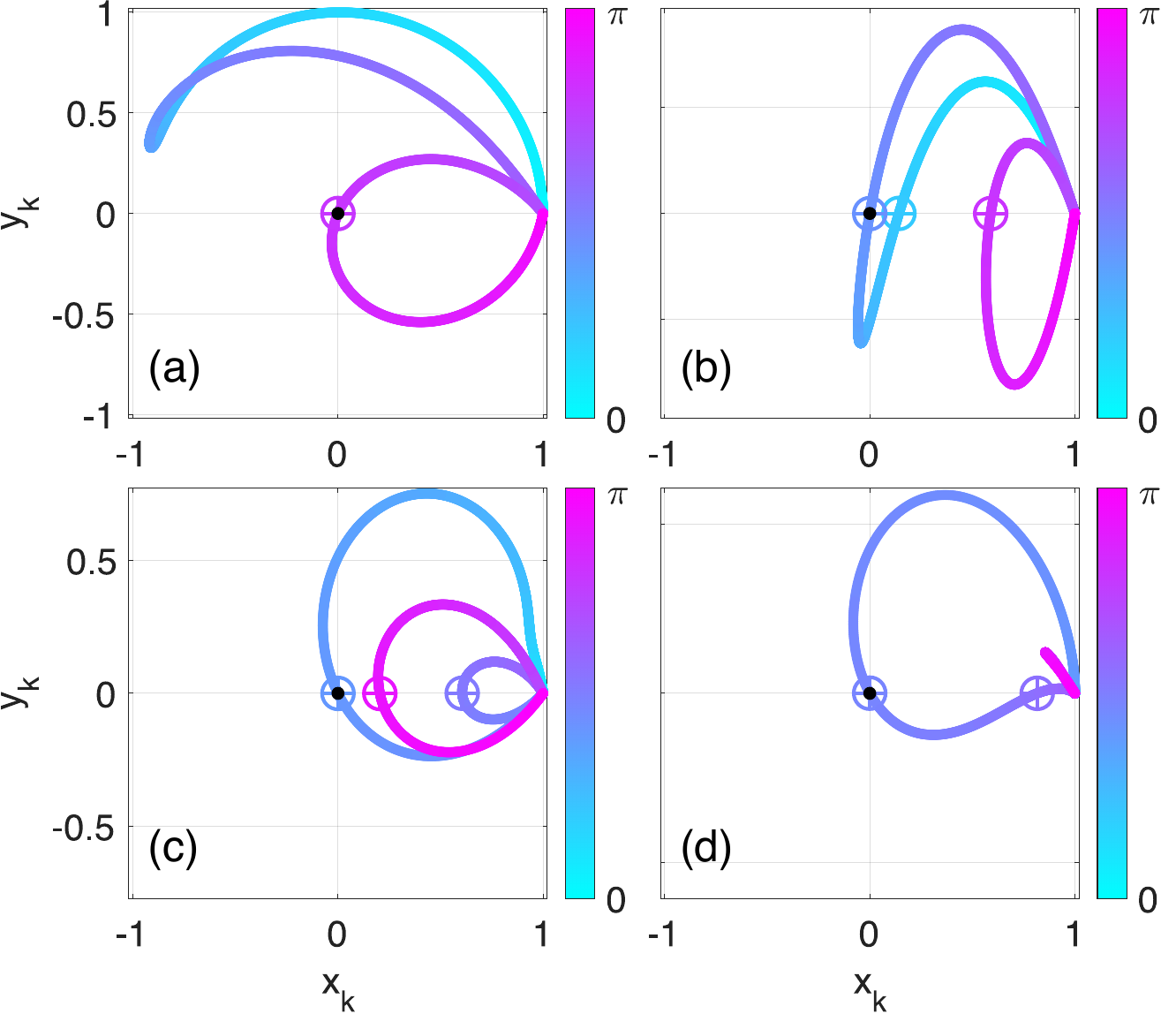}
    \caption{Trajectories of $\vec{r}_k$'s for the same quenches and system size in Fig. \ref{fig3_LR_1ks3ks}. Time is frozen at the first critical time $t_1^*$. Black dots indicate the origin. The ``$\oplus$" signs pin the corresponding  critical momenta.}
    \label{fig5_rk_1ks3ks}
\end{figure}

In the original SSH model, there are only two phases, the trivial $W_0$ phase and the topological $W_1$ phase, with $\eta$ as the driving parameter. The condition for a DQPT is determined by the equation $\cos k^* = -( 1 + \eta_i\eta_f ) / ( 1 - \eta_i\eta_f )$. Given the quantum critical point at $\eta_c = 0$, quenching the original SSH model guarantees at most one critical momentum. On the other hand, the addition of $t_c,t_d$ hoppings in the extended model introduces two additional topological phases -- the $W_2$ and $W_{-1}$ phases, and one can have DQPTs with at most three critical momenta.

Figure \ref{fig:Dphasediagram} shows the dynamical phase diagram defined by the number of critical momenta $n_{k^*}$ for an initial state taken from each of the four equilibrium phases. Interestingly, for the quench between $W_0$ and $W_1$ phases, a dynamical phase boundary separating the $1k^*$ (the regular) and $3k^*$ (the anomalous) regions within the same equilibrium phase is observed (Figs. \ref{fig:Dphasediagram}(a) and \ref{fig:Dphasediagram}(b)). We also find that DQPTs can occur for quenches without crossing an equilibrium phase boundary. Especially for quenches within the $W_2$ or $W_{-1}$ phases (Figs. \ref{fig:Dphasediagram}(c) and \ref{fig:Dphasediagram}(d)), there exists a finite region in the dynamical phase diagram that DQPTs take place within the same phase.  
\red{These observations motivate us to define the \textit{type-B anomalous} DQPT as the regions where $n_{k^*}$ exceeds $|\Delta W|$. In later parts we will elucidate theoretically that all these type-B anomalous DQPTs originate from the same cause.} 


The Loschmidt rate (LR) and DTOP for some selected quench cases are shown in Fig. \ref{fig3_LR_1ks3ks}. It clearly shows the difference for quenches to the anomalous regions and the regular regions. Namely, there is one non-analytical peak in LR associated with the upward integer jump on DTOP in Fig. 
\ref{fig3_LR_1ks3ks}(a), whereas in Fig. \ref{fig3_LR_1ks3ks}(b) two extra critical times result when quenched to the $3k^*$ region for $W_1 
\rightarrow W_0$. On the other hand, Fig. \ref{fig3_LR_1ks3ks}(c) presents the quench with the three $k^*$'s, the highest possible number of critical momenta, from $W_2$ to $W_{-1}$ phase, where three kinks at distinct times can be found. Figure \ref{fig3_LR_1ks3ks}(d) shows the DQPT for quench within the lower $W_2$ phase and we clearly see two finite critical times with $\nu_D(t)$ jumped up at $t_1^*$ and down at $t_2^*$.

Here, we define the regions with $n_{k^*}$ exceeding $|\Delta W|$ as the type-B anomalous regions and present the possible cause and ways to distinguish them from regular DQPTs. Note that in both anomalous cases presented in Table \ref{tab:sum}, two additional critical momenta are found \red{(comparing column 1 and 3, and column 2 and 4)}. The two extra $k^*$'s are originated from a complex conjugate pair of $k^*$ in $\vec{d}_i(k^*) \cdot \vec{d}_f(k^*)=0$, where they share the same real part. As the postquenched parameters enter the anomalous region (the $3k^*$ region for quench between $W_0$ and $W_1$, and the $2k^*$ region for quench within the $W_{-1}$ or $W_2$ phases), the imaginary part of the two critical momenta vanishes and they transform to the real critical momenta, giving rise to extra real critical times. \red{This ensures that the parity of $n_{k^*}$ is conserved in both regions. 
For example, a $W_2\rightarrow W_2$ quench gives either zero $k^*$ or 2 $k^*$'s, while a $W_1\rightarrow W_0$ quench gives either 1 $k^*$ or 3 $k^*$'s.}

\red{The emergence of the complex-conjugate pair is visualized in Fig. \ref{fig4_ks_traj}, in which it} illustrates the locations of $k^*$'s on the complex plane around the dynamical phase boundary. The complex conjugate pairs zone in the real axis when the final quench parameters approach the dynamical phase boundaries. \red{The two critical momenta overlap} right at the dynamical phase boundary, implying only one additional critical time resulted. \red{Notice, however, that this is to be differentiated from the degenerate critical modes observed in quenched clustered Ising model, which is related to the structure of its energy spectrum \cite{typeBanomalous}.} When quenched further into the anomalous regions, the two critical momenta separate farther apart from each other on the real axis and two distinct critical times emerge.

\red{Apart from the location of $k^*$'s,} the vector $\vec{r}_k$ also reveals distinct features between regular and type-B anomalous regions, \green{as summarized in Table \ref{tab:sum}. Figure \ref{fig5_rk_1ks3ks} shows the trajectories of $\vec{r}_k$ for the respective quench cases in Fig. \ref{fig3_LR_1ks3ks} at the first critical time.} For the regular case where $n_{k^*} = |\Delta W|$, each $k^*$ is ``assigned" a separate loop in the trajectory of $\vec{r}_k$. It is clearly seen in Fig. \ref{fig5_rk_1ks3ks}(a), quenching from $W_1$ phase to $1k^*$ region in $W_0$ phase where we have one loop for one critical momentum, and Fig. \ref{fig5_rk_1ks3ks}(c), quenching from the bottom $W_2$ phase to the left $W_{-1}$ phase in the phase diagram shown where we have three loops, each containing one $k^*$. On the other hand, for the anomalous case with quench to $3k^*$ in Fig. \ref{fig5_rk_1ks3ks}(b) and the quench within the bottom $W_2$ phase in Fig. \ref{fig5_rk_1ks3ks}(d), the critical momentum pair gathers in the same loop instead of staying in two separate loops, which further suggests the same origin of the two extra critical momenta. This feature still holds true for the SSH model with even longer range hoppings as discussed in Appendix \ref{sec:app_A}.
One thing to notice is the extra loop for quench $W_1 \rightarrow W_0$ to $1k^*$ region in Fig. \ref{fig5_rk_1ks3ks}(a) that has no critical momentum contained in it. This extra loop corresponds to the imaginary critical momentum pair that are not yet landed onto the real axis and cosine range. For the postquenched Hamiltonian approaching the dynamical boundary, the loop will eventually touch the origin, marking the emergence of the $k^*$ pair.

In the above discussions and those that follow, we focus on the type-B anomalous DQPTs. It is worth noting that type-A anomalous DQPTs are also observed in this system. The results and the corresponding discussions are presented in Appendix \ref{sec:app_B}.


\section{Entanglement Entropies}
\label{sec:EE}


In this section, we consider the time evolution of entanglement in different quench cases. We only focus on the short-time behavior of the entanglement entropies (EE) near DQPT. Our results suggest the EE signals the DQPT by its extremum, which is usually subsystem dependent except for quenches deep inside the phase where the dynamics is quite coherent for all the subsystems. The entanglement is quantified by the von-Neumann entropy defined as
\begin{equation}
\label{eqn:entangl}
    S_\Omega = -\text{Tr}(\rho_{\Omega} \log \rho_{\Omega}),
\end{equation}
where $\rho_{\Omega}$ is the reduced density matrix of the subsystem $\Omega$. 

In non-interacting systems, the 
entanglement entropy can be calculated by 
\begin{equation}
\label{eqn:calculateEntangle}
    S_\Omega = -\sum_{j}[ \xi_j \ln{\xi_j} + ( 1 - \xi_j )\ln( 1 - \xi_j ) ], 
\end{equation}
where $0\le\xi_j\le 1$ are the eigenvalues of the correlation matrix $C_{\mu\nu}=\braket{\hat{c}_\mu^{\dagger} \hat{c}_\nu}$ for $\mu,\nu\in \Omega$ \cite{Poyhonen2021,Latorre2009,Zhang2014,Peschel2003}.
The eigenvalues of the correlation matrix measure the probabilities that the states with energies $\ln[(1-\xi_j)/\xi_j]$ in the single-particle entanglement spectrum are occupied \cite{Wybo2021}. 

Equation (\ref{eqn:calculateEntangle}) can also be viewed as the Shannon entropy of a Bernoulli distribution among the eigen-levels of the correlation matrix \cite{Zhang2014}. The largest contribution to the entanglement entropy comes from an eigenvalue of $\xi_j = 0.5$, meaning the eigen-level is simultaneously occupied and unoccupied. 
In the original SSH model under the periodic boundary condition, the correlation matrix spectrum is doubly degenerated at $\xi_j = 0.5$ in the topological phase, which corresponds to the boundary modes at two ends of the subsystem. 
In the quasiparticle picture, this induces a position uncertainty in the occupation which further translates to the momentum uncertainty, leading to an increase of entropy \cite{Wybo2021}. In the following, the time evolution of EE will be revealed in the vicinity of DQPTs, followed by further investigations on the correlation matrix spectrum in Sec. \ref{sec:CMS}.

\subsubsection{EE in the original SSH model}

\begin{figure}
    \centering
    \includegraphics[width=8cm]{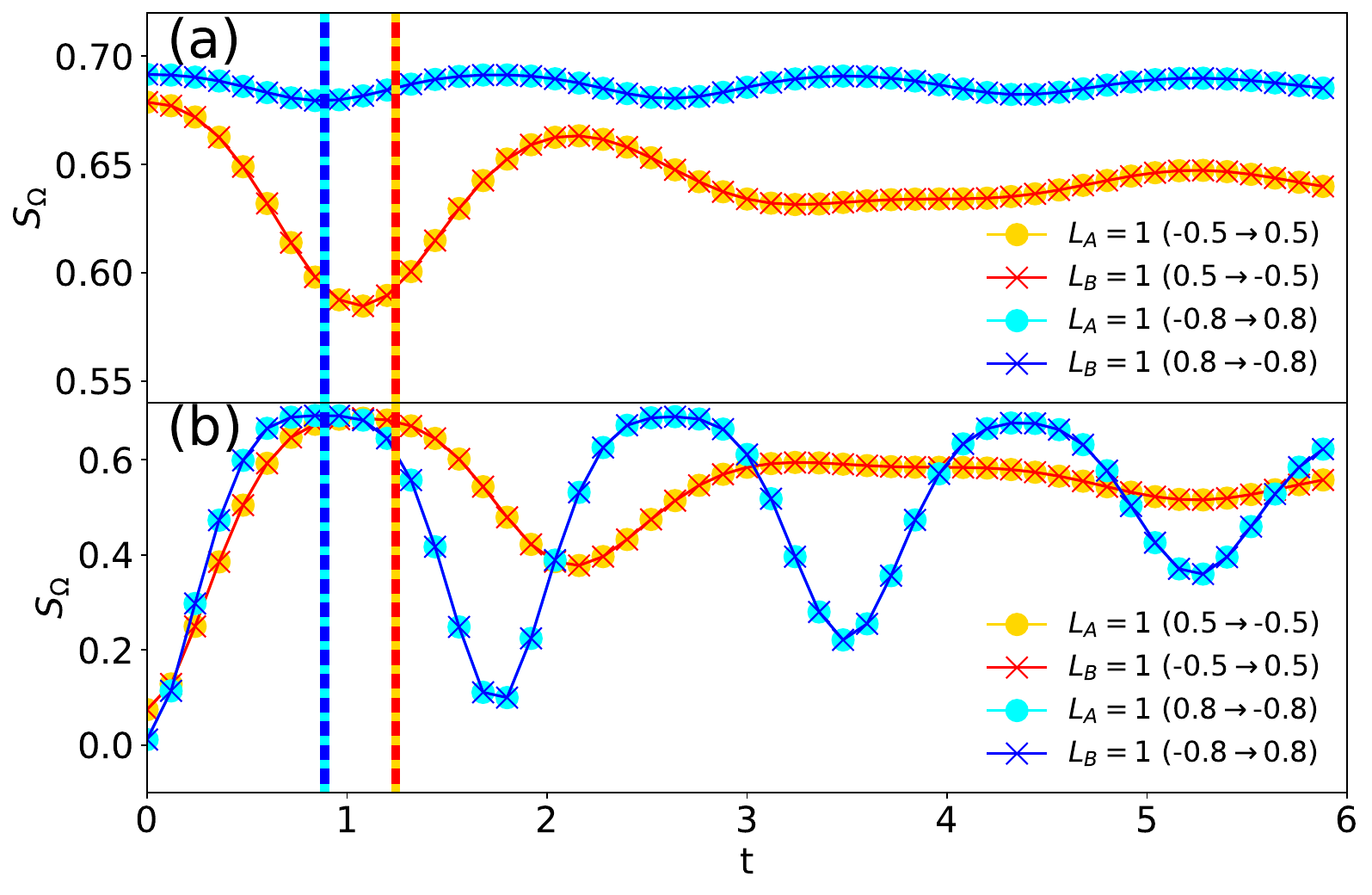}
    \caption{Time evolution of the entanglement entropy in the original SSH model ($t_c=t_d=0$) under a symmetric quench. The legend indicates the subsystem considered and the quench parameters $(\eta_i\rightarrow\eta_f)$. 
    Here $L=80$ and only the first critical time is shown.}
    \label{fig:oriSSH}
\end{figure}

We start from the original SSH model in equilibrium. Two premises are stated below in order to facilitate discussions: 1. The $W_0$ phase favors stronger $t_a$ hoppings ($\eta>0$) while the $W_1$ phase favors stronger $t_b$ hoppings ($\eta<0$) \cite{yu2016,Sacramento2023}; and  2. physically, EE reflects the information lack on the whole system when one only considers a certain subsystem. In other words, higher entropy corresponds to a higher lack of knowledge in the subsystem and vice versa. Combining the above statements, we argue that the equilibrium EE of $L_A=1\red{/L_B=1}$ (refer to Figs. \ref{fig:chain}(a) \red{and \ref{fig:chain}(b)}) should be larger in the $W_1\red{/W_0}$ phase than in the $W_0\red{/W_1}$ phase. 
On one hand, the system favors $t_a\red{/t_b}$ over $t_b\red{/t_a}$ hoppings in the $W_0\red{/W_1}$ phase. On the other hand, $L_A=1\red{/L_B=1}$ contains a $t_a\red{/t_b}$ bond but not a $t_b\red{/t_a}$ bond. Therefore, $L_A=1\red{/L_B=1}$ contains more information about the $W_0\red{/W_1}$ phase, resulting in a smaller value of EE at equilibrium \red{(see Fig. \ref{fig:oriSSH}(b) $t=0$)}. Similarly, the $W_1\red{/W_0}$ phase favors $t_b\red{/t_a}$ over $t_a\red{/t_b}$ hoppings, resulting in a higher EE for the subsystem $L_A=1\red{/L_B=1}$ due to its lack of information \red{(see Fig. \ref{fig:oriSSH}(a) $t=0$)}. 
These are especially true for the blue curves, as they demonstrate quenches deeper inside the corresponding phases. 
\red{Therefore, in the original SSH model, the EE can be large or small in both the $W_0$ and the $W_1$ phases} \green{depending on the choice of the subsystem.}

Now we move on to the discussion of the behavior near DQPT for the original SSH model. \red{Here, our main focus is on the EE of small subsystems. For the half-block EE, there is an interesting discussion in Ref. \cite{Sedlmayr2018}, where the authors demonstrated the relation between the time evolution of the two-point correlations and the oscillatory behaviour in EE. }In Fig. \ref{fig:oriSSH}(a), a local minimum in EE for subsystem $L_A=1$ is attained around DQPT for quench $W_1\rightarrow W_0$ phase (orange and light blue curves). Following the discussion above, this implies an influx of information from the surroundings to the subsystem $L_A=1$ in the vicinity of DQPT induced by the quench. Figure \ref{fig:oriSSH}(b) shows quenches where EEs achieve local maxima near DQPT. The maximal behavior of EE suggests an outflow of information from the corresponding subsystem to the surroundings in the vicinity of DQPT. One can also notice the exact opposite happened to the subsystem $L_B=1$, where the EE is maximum and minimum for $W_1\rightarrow W_0$ and $W_0\rightarrow W_1$, respectively, around the critical time. \red{We can understand this as the "role" of $W_0$ and $W_1$ can be switched by choosing an appropriate subsystem.}

The above observations motivate us to introduce a protocol of choosing a suitable subsystem that can realise minimum/maximum EE near DQPT. Let us take the case of having a minimum EE as an example. First, noting the phase in which the initial state is taken from, the subsystem has to be chosen such that the corresponding favored hopping bond is not completely contained inside the subsystem. For instance, in order to realize minimum near DQPT for the quench $W_1\rightarrow W_0$, one notices that the $W_1$ is the initial phase which favors $t_b$ hoppings. The protocol above suggests us to choose a subsystem that does not contain a complete $t_b$ bond, thus $L_A=1$ satisfies this condition. The EE for the corresponding subsystem achieves a local minimum near DQPT when one activates a quench from $W_1\rightarrow W_0$ phase. On the other hand, to realize maximum in EE around DQPT, a suitable choice of the subsystem will be the one which contains a complete bond of the favored hopping of the initial phase. 


\subsubsection{EE in the extended SSH model}
In the dynamical phase diagrams discussed in Sec. \ref{sec:ks_diag}, we focused on the case of $\eta=0$ in the extended SSH. This corresponds to where the $t_a$ and $t_b$ hoppings are equally weighted. Surprisingly, the $W_0$ and $W_1$ phases still emerge in the equilibrium phase diagram in the regime where $t_c \approx -t_d$ (Fig. \ref{fig:Dphasediagram}(a)). This suggests although $t_a$ and $t_b$ are equally weighted, the system can still favor $W_0$ and $W_1$ phase if $t_c$ and $-t_d$ has approximately equal weight. 

\begin{figure}[t]
    \centering
    \includegraphics[width=8.5cm]{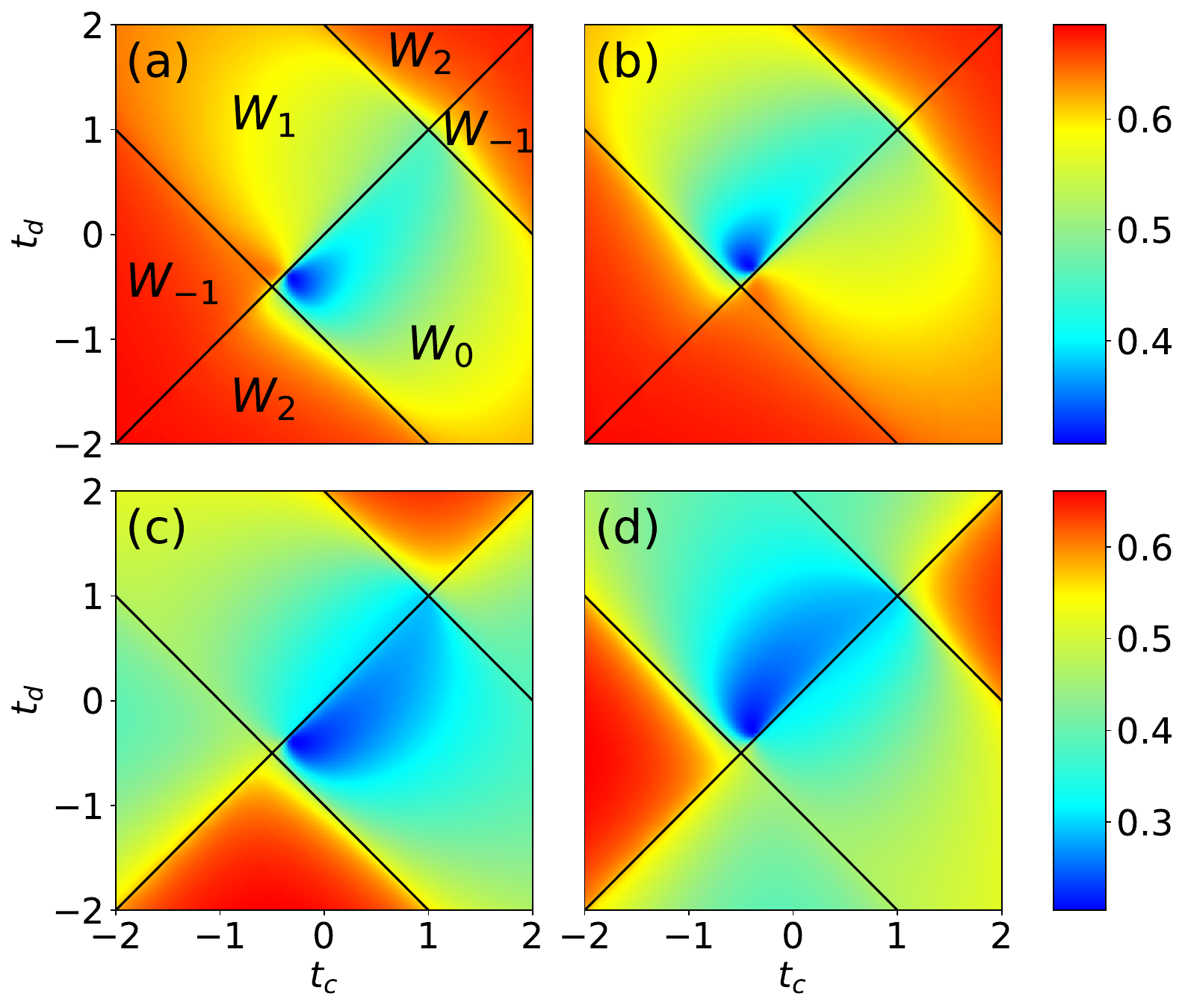}
    \caption{Entanglement entropy per unit cell with various subsystems (a) $L_A=1$, (b) $L_B=1$, (c) $L_A=2$, and (d) $L_B=2$ in the equilibrium extended SSH model. The top color bars are meant for (a) and (b), while the bottom ones are meant for (c) and (d). \green{The black solid lines indicate the equilibrium phase boundary.} The simulation was performed with $L=120$.}
    \label{fig:phasediagram}
\end{figure}

We begin the discussion of the extended SSH model by first considering the EE in its equilibrium phase diagram shown in Fig. \ref{fig:phasediagram}. Recall that EE measures the information lack, as such it enables us to quantify the degree of non-locality for each phase. The figure suggests that the ground states of the $W_2$ and $W_{-1}$ phases are both more non-local than that of the $W_0$ and $W_1$ phases since most regions in $W_2$ and $W_{-1}$ phases have higher or approximately equal entropy values than those of $W_0$ and $W_1$ phases, regardless of the subsystems we choose. Therefore, it is natural to separate $W_2$ and $W_{-1}$ as one class of phase, while the other two being another class of phase. Another reason why we propose these two different classes is due to symmetry. If one performs the following mapping $(t_c,t_d,L_A=\alpha)\mapsto(t_d,t_c,L_B=\alpha)$, where $\alpha\in\mathbb{N}$, the equilibrium EEs remain unchanged. In another perspective, the EEs for $L_A=\alpha$ and $L_B=\alpha$ differ only by a reflection along the $t_d=t_c$ diagonal in Fig. \ref{fig:phasediagram}. Therefore, we suggest categorizing $W_0,W_1$ as class-1 phases, and $W_2,W_{-1}$ as class-2 phases, where phases from class 2 have higher degree of non-locality than those in class 1. Phases in the same class have equal degree of non-locality. \green{Such a classification of the phases is necessary to extract the generic features of EE around DQPT. This may serve as an alternative approach to probe entanglement physics for DQPTs with $n_{k^*} > 1$ which was previously found with no apparent connection between EE and DQPTs via conventional approach \cite{multibandSSHDQPT}.} We will demonstrate quenches between or within classes have distinctive EE properties near DQPT in the following discussion.

\begin{table*}[]
\begin{tabular}{|l|l|l|}\hline 
                         & \hspace{40 pt} Class 1                 & \hspace{50 pt} Class 2                          \\
                         \hline
\multirow{2}{*}{Class 1} & $n_{k^*}=|\Delta W|$
: quench dependent                 & \multirow{2}{*}{ $n_{k^*}=|\Delta W|$
: max for deep quench}          \\
                         & $n_{k^*}>|\Delta W|$: max/min                &                                  \\
                         \hline
\multirow{2}{*}{Class 2} & \multirow{2}{*}{ $n_{k^*}=|\Delta W|$
: always min} & $n_{k^*}=|\Delta W|$
: max for deep quench \\
                         &                         & $n_{k^*}>|\Delta W|$: always min  \\ 
                         \hline
\end{tabular}
\caption{Summary of the properties of EE of a suitably chosen subsystem (see text for details) near DQPT in the extended SSH model. Class 1 contains $W_0$ and $W_1$ phases, while class 2 consists of $W_2$ and $W_{-1}$ phases. The diagonal cells represent quenches within the same class. For example, $W_2\rightarrow W_2$ and $W_{-1}\rightarrow W_2$ belong to the second diagonal cell in the table, while $W_1\rightarrow W_0$ belongs to the first diagonal cell. The off-diagonal cells refer to quenches between class 1 and class 2. Class $2 \rightarrow 1$ corresponds to the (2,1) cell, while class $1 \rightarrow 2$ is represented by the (1,2) cell. An example of class $2 \rightarrow 1$ quench is $W_2\rightarrow W_0$.}
\label{tab:ent}
\end{table*}

The results in the original SSH suggest that the extremum properties of EE near DQPT can be manipulated by choosing a suitable subsystem. However, due to the richness of phases in the extended SSH model and the absence of the notion of classes of phases in the original SSH, the protocol of choosing the subsystem introduced in the original SSH model may not hold in some cases. Table \ref{tab:ent} summarizes the \red{generic} findings in quenches among the classes, and the corresponding details are discussed below. 

\paragraph{Class 1 $\rightarrow$ Class 1:}

\begin{figure}[t]
    \centering
    \includegraphics[width=8cm]{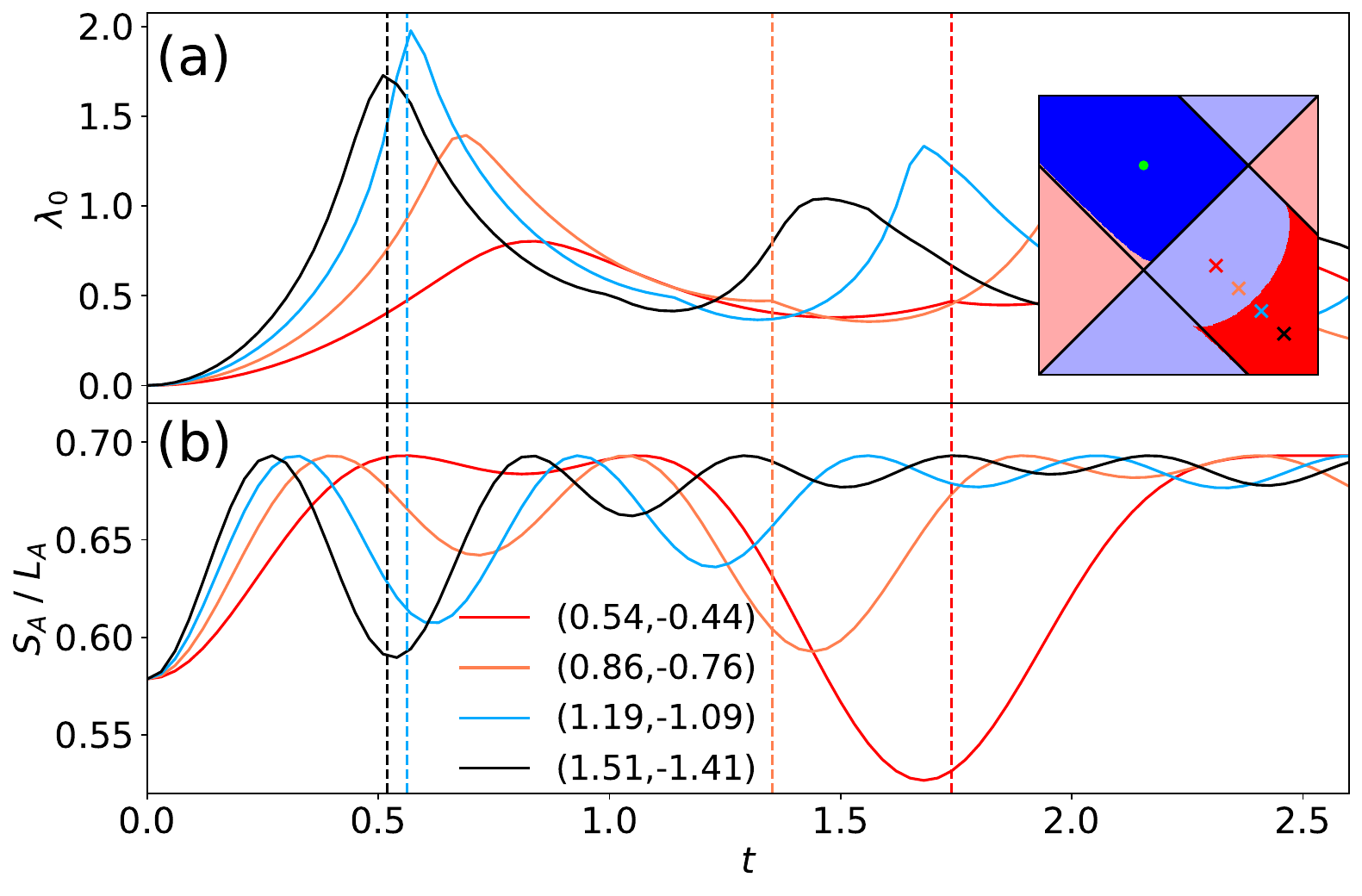}
    \caption{ (a) Loschmidt rate functions and (b) time evolution of the EEs with subsystem $L_A=1$ for quenches $W_1 \rightarrow W_0$ of the extended SSH model. The initial parameter is $(t_c^i,t_d^i)=(-0.5,1)$, quenching to the various final parameters $(t_c^f,t_d^f)$ that lies on the path $t_d^f=-t_c^f+0.1$. The exact value of final parameters are shown in the legend. Colored dashed lines indicate the first critical time for the corresponding quench cases. Here $L=80$. Initial and final points are indicated as green dot and crosses respectively in the inset color map of (a).}
    \label{fig:ent1-0}
\end{figure}

The first diagonal block in Table \ref{tab:ent} corresponds to quenches within phases from class 1. We notice that the behavior of EE in the vicinity of DQPT is mostly similar to that of the original SSH model, but with certain additional features. The quenches involving the $3k^*$ anomalous region, i.e. $n_{k^*}>|\Delta W|$, have similar features as the original quenched SSH model. For example, Fig. \ref{fig:ent1-0} shows the time evolution of EE for $L_A=1$ along some selected quench paths from $W_1$ to $W_0$ phases, enclosing a dynamical phase boundary where $n_{k^*}$ changes within the same equilibrium phase. The EE in the $3k^*$ region 
exhibits a local minimum near DQPT for $L_A=1$ (see Fig. \ref{fig:ent1-0}(b)). 

For the $1k^*$ regular region, i.e. $n_{k^*}>|\Delta W|$, the behavior of the entanglement around DQPT is, in general, quench dependent. The EE may be a minimum, a maximum, or neither of them around DQPT. The inconsistency is due to the fact that the quench is near the equilibrium phase boundary in the $1k^*$ regime, as compared to the $3k^*$ regime where the quench is well away from the $t_d = t_c$ phase boundary. Quenches to places near the equilibrium phase boundary sometimes may not yield consistent results due to the rapid change of properties of the systems. These observations may suggest that the introduction of $t_c$ and $t_d$ hoppings and equalizing the weight between $t_a$ and $t_b$ hopping have the effect of recovering the EE property near DQPT of the original SSH model only in the $3k^*$ region, but partially in the $1k^*$ region. 
%

There is an interesting feature regarding quenches to regular and anomalous region in the extended SSH model. Figure \ref{fig:ent1-0}(b) shows the DQPT occurs at the second local minimum (sometimes the third or fourth and so on) of the EE for the $1k^*$ region, whereas for the $3k^*$ region, DQPT consistently occurs at the first local minimum of EE. In fact, we always find the DQPT in the regular region occurs at a later time, while DQPTs in anomalous region happen earlier. This echoes with Table \ref{tab:sum} that regular regions often have longer $t^*$, while anomalous region has shorter ones. This often results in a finite gap in the critical time when one quenches along a path from the regular region to the anomalous region, for example the cases shown in Fig. \ref{fig:ent1-0}.  One may interpret the first minimum in the EE in the $1k^*$ region as a pre-development of a DQPT. Once the quench enters the $3k^*$ region, DQPT emerges at the first minimum immediately.

\begin{figure}[t]
    \centering
    \includegraphics[width=8.5cm]{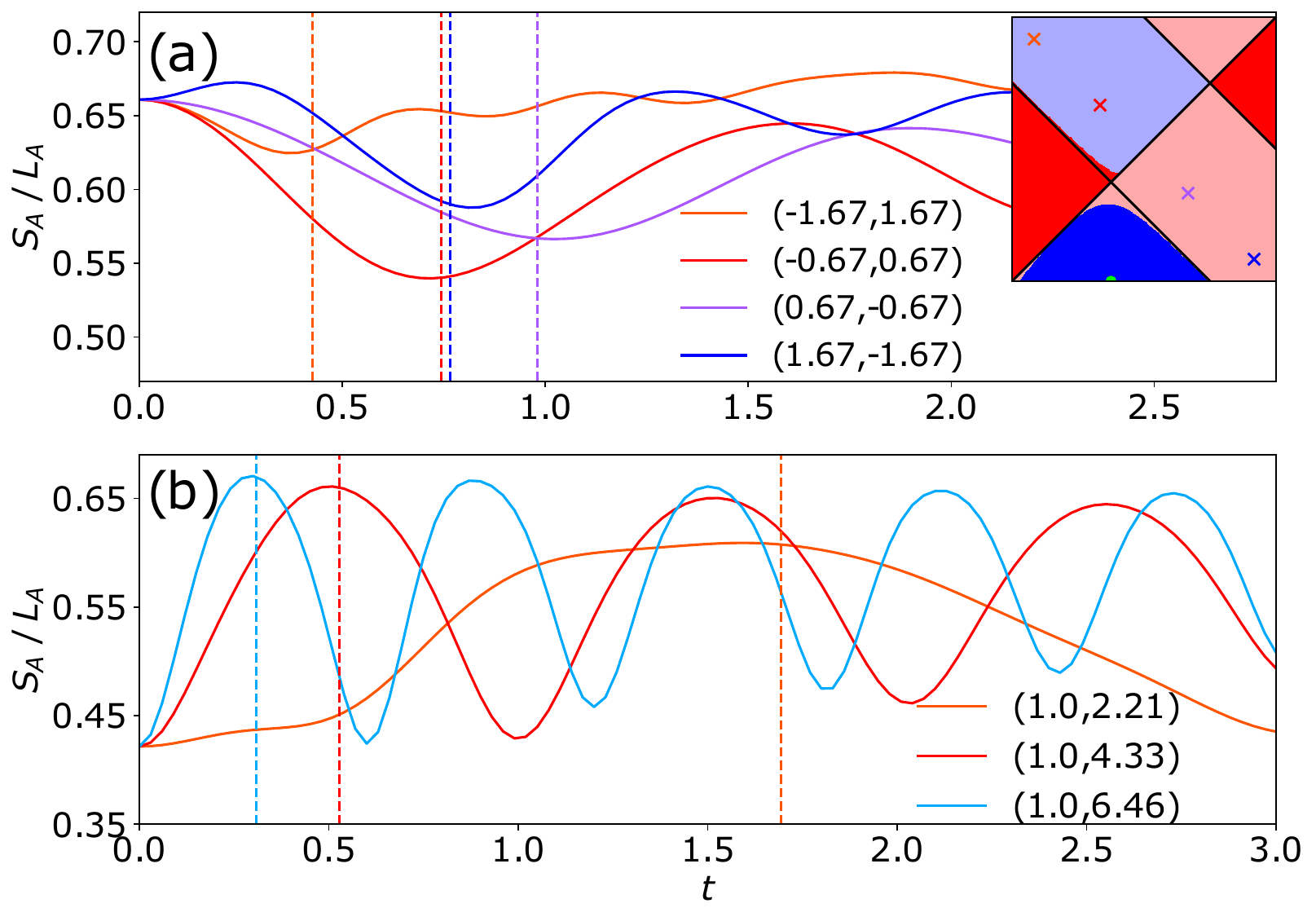}
    \caption{Time evolution of the entanglement entropies per unit cell with subsystem $L_A=2$ involving quenches between the two classes: (a) from class 2 to class 1 and (b) from class 1 to class 2. The initial point in (a) and (b) are $(t_c^i,t_d^i)=(-0.5,-2)$ and $(t_c^i,t_d^i)=(-0.5,1)$ respectively. Legends denote the final point $(t_c^f,t_d^f)$ of the quenches, tracing a path of (a) $t_c^f=-t_d^f$ and (b) $t_c^f=1$. The vertical dashed lines pin the first critical time for each of the quench cases. Here $L=80$. Initial and final points of quenches in (a) are indicated as green dot and crosses respectively in the inset panel.}
    \label{fig:ent-betweenClass}
\end{figure}

\begin{figure}[t]
    \centering
    \includegraphics[width=8cm]{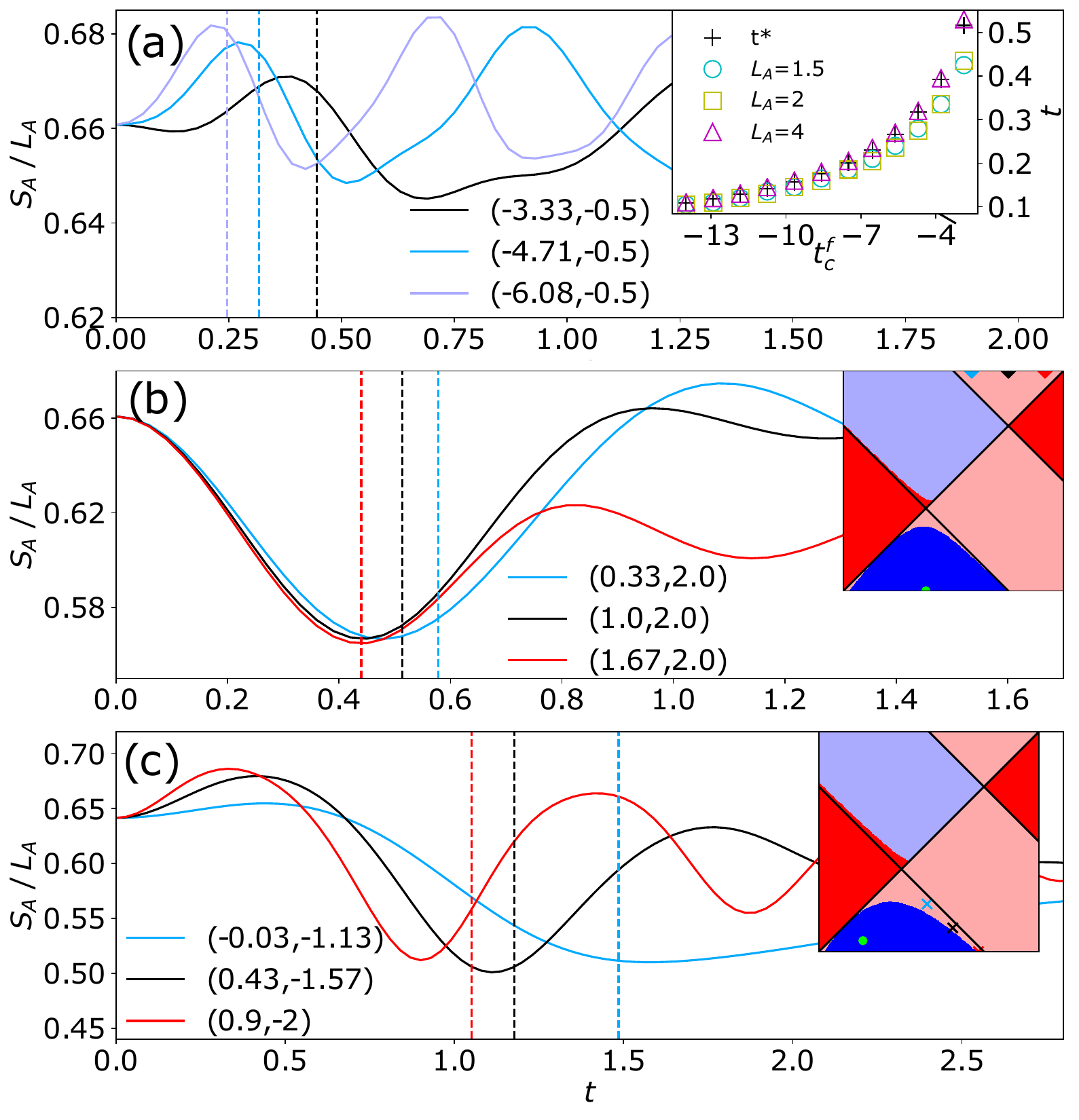}
    \caption{Time evolution of the entanglement entropies per unit cells for quenches within class 2. (a) From $(t_c^i,t_d^i)=(-0.5,-2)$ along the path $t_d^f=-0.5$, (b) $(t_c^i,t_d^i)=(-0.5,-2)$ along the path $t_d^f=2$ and (c) $(t_c^i,t_d^i)=(-1.2,-1.8)$ along the path $t_d^f \approx -0.93t_c^f-1.16$. All subsystems are chosen to be $L_A=2$. The precise final quench parameter $(t_c^f,t_d^f)$ is indicated in the legends. Inset in (a) shows the critical times and the times when the EE for various subsystems achieves a maximum as a function of $t_c^f$. The colored dashed lines show the first critical time for the corresponding quench case. Here $L=80$. Inset color maps in (b) and (c) illustrate the quench cases, where final parameters are indicated as diamonds and crosses respectively while the initial points are indicated as green dots.}
    \label{fig:ent-2and-1}
\end{figure}

\paragraph{Class 1 $\leftrightarrow$ Class 2:} 

We now discuss the result presented in the off-diagonal block in Table \ref{tab:ent}. Analogous to original SSH model, where $W_0$ and $W_1$ phase favors $t_a$ and $t_b$ hoppings respectively, we also need to determine the favorable hopping corresponding to $W_2$ and $W_{-1}$ phases before proceeding the discussion. The equilibrium phase diagram in Fig. \ref{fig:phasediagram}(a) suggests that $W_2$ and $W_{-1}$ phases are highly influenced by the $t_d$ and $t_c$ hopping terms respectively. Therefore, we propose $W_2$ and $W_{-1}$ phase favors $t_d$ and $t_c$ hoppings respectively. 


Figure \ref{fig:ent-betweenClass} shows the time evolution of EE quenching from class-1/2 to class-2/1 phases.  Figure \ref{fig:ent-betweenClass}(a), in particular, shows several examples of quenches from class 2 to 1 (which corresponds to the (2,1) block in Table \ref{tab:ent}). The protocol of producing a EE minimum in the vicinity of DQPT still holds, but not for the maximum. According to the protocol in the previous section, the suitable subsystem for the initial phase $W_2$ and $W_{-1}$ is $L_A=2$ and $L_B=2$ respectively, where the corresponding favored hoppings are not completely contained (see Fig. \ref{fig:chain}). The result suggests quenches from a more non-local phase to a less non-local phase can result in information gain in the subsystem in the vicinity of DQPT, as long as a suitable subsystem is chosen. 

Figure \ref{fig:ent-betweenClass}(b) displays quenches from class 1 to class 2 (which corresponds to the (1,2) block in Table \ref{tab:ent}). On one hand, unfortunately, no analogy can be made from the protocol of inducing maximum/minimum near DQPT if one considers shallow phase quench. Deep-phase quench here refers to the situation where the postquenched Hamiltonian is dominated by one of the hoppings $t_a$, $t_b$, $t_c$ or $t_d$ over the rest. For example, an extremely large value of $t_d$ and small values of $t_a$, $t_b$ and $t_c$ is considered as deep inside $W_2$ phase. On the other hand, if one considers quenches involving the final parameter that is sufficiently deep inside the phase, it is possible to realize consistent EE maximum near DQPT for all the subsystems, even for the half-block case (not shown here). 


\paragraph{Class 2 $\rightarrow$ Class 2:}
Figure \ref{fig:ent-2and-1} shows quenches between phases in class 2, which correspond to the second diagonal cell in Table \ref{tab:ent}. We first consider the case where $|\Delta W| \neq 0$. Similar to the case of class 1 quench to class 2, there is no general way to always attain an extremum if we use the protocol discussed concerning the original SSH model. The only consistent way to attain maximum is to quench sufficiently deep inside the phase. Figure \ref{fig:ent-2and-1}(a) shows an example of deep-phase quench from $W_2$ to $W_{-1}$ phase on the left. The inset shows the first critical times (the plus markers) and the times at which EE for various subsystem attains maximum (the hollow symbols) as a function of $t_c^f$. It indicates the deeper we quench inside the phase (more negative $t_c^f$), the more the maximum EE and the critical time are aligned. Similar results are obtained for the quench from $W_{-1}$ to $W_2$ (not shown here). 
\green{Note that the result holds for any subsystems including the case of a half-block (refer to the right panel of Fig. \ref{fig:CMS_W2ntoWn1n_deep}, which shows the time evolution of the half-block correlation matrix spectrum for the quench $W_2\rightarrow W_{-1}$. The eigen-levels tend towards the Fermi level around DQPT results in higher EE. More detail will be discussed in the next section).} 

One can realize another type of quenches that stay within \red{class-2 phases}. Unlike all the quenches we mentioned above, this quench has $n_{k^*}>\Delta W=0$, which is a type-B anomalous DQPT. Figures \ref{fig:ent-2and-1}(b) and \ref{fig:ent-2and-1}(c) show quenches between the two $W_2$ phases, with the former quench across the equilibrium phase boundary, while the latter does not. Using the protocol discussed in the original SSH model to choose the appropriate subsystem ($L_A=2$), these two varieties of quenches attain a local minimum in the vicinity of DQPT. Similar result can be observed if one changes the subsystem to $L_B=2$ for quenches within the $W_{-1}$ phase. Recall that the EE  represents the information lack in a particular subsystem with respect to the whole system. 
 The result suggests that in the vicinity of DQPT, the subsystem with 2 unit cells contain more information about the whole system as compared to $t=0$. A possible application is that one may be able to extract significant amount of the whole system's information by just performing measurements on a small subsystem around DQPT of this type, without probing larger subsystems. 

In \red{Appendix \ref{sec:app_A1}, the model is extended to include even longer range of hoppings and the time evolution of the EE for some quench cases are presented. The results suggest the generalizability of our findings. }

\section{Correlation Matrix Spectra}
\label{sec:CMS}

\begin{figure}
    \centering
    \includegraphics[width=8cm]{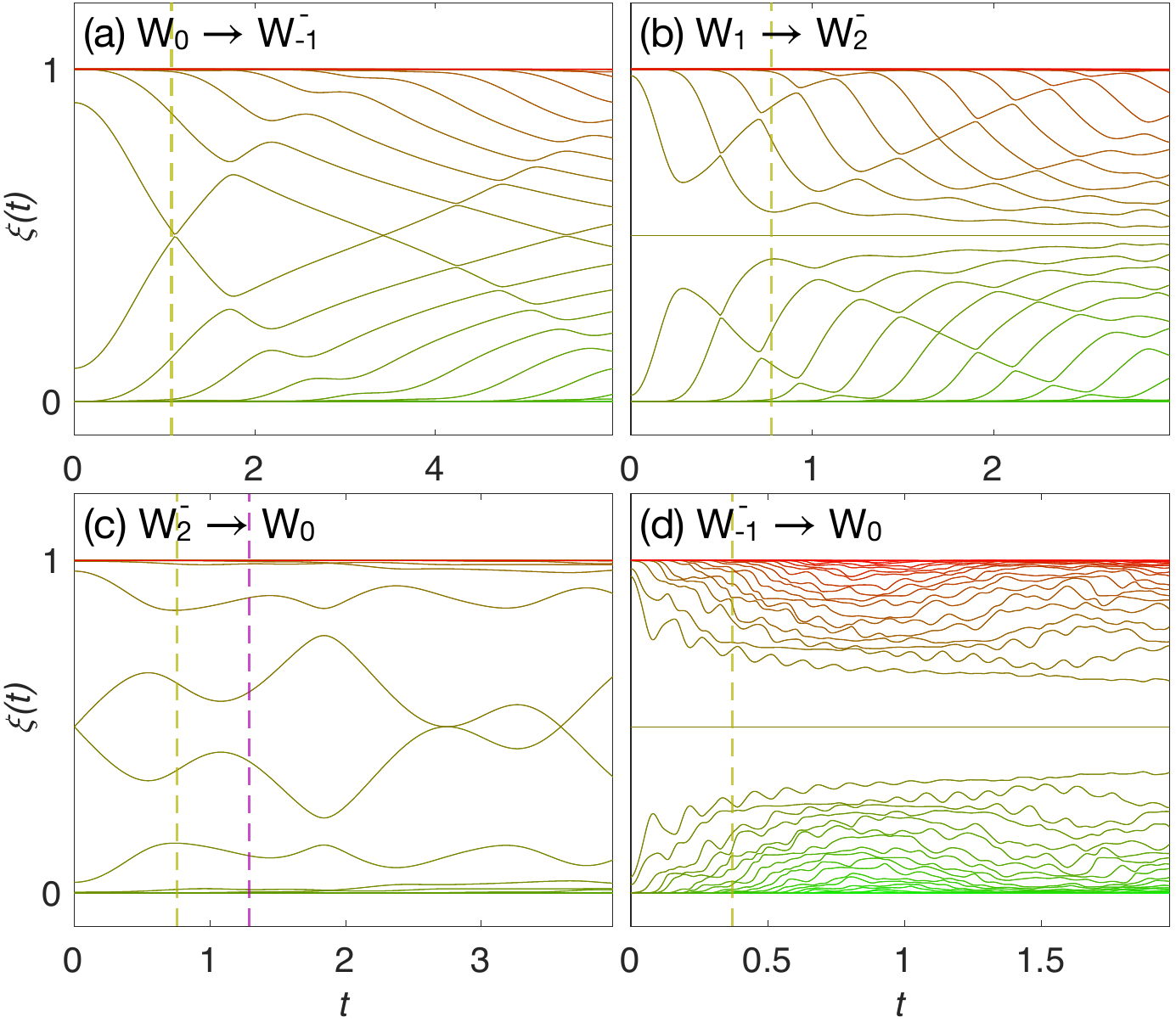}
    \caption{Time evolution of correlation matrix spectrum of a half-block subsystem for quenches (a) $(1,-0.5) \rightarrow (-1.7,-0.5)$, (b) $(-0.5,1) \rightarrow (-0.11,-1.7)$, (c) $(-1.2,-1.8) \rightarrow (0.9,-0.3)$ and (d) $(-1.8,-1.2) \rightarrow (2,-1.9)$ for an $L = 80$ system. The prequenched parameters are the same as those in Fig. \ref{fig:Dphasediagram}.  Colored dashed lines indicate the first appearance of the critical times.}
    \label{fig:CMS_phases}
\end{figure}

\begin{figure*}
    \centering
    \includegraphics[width=16cm]{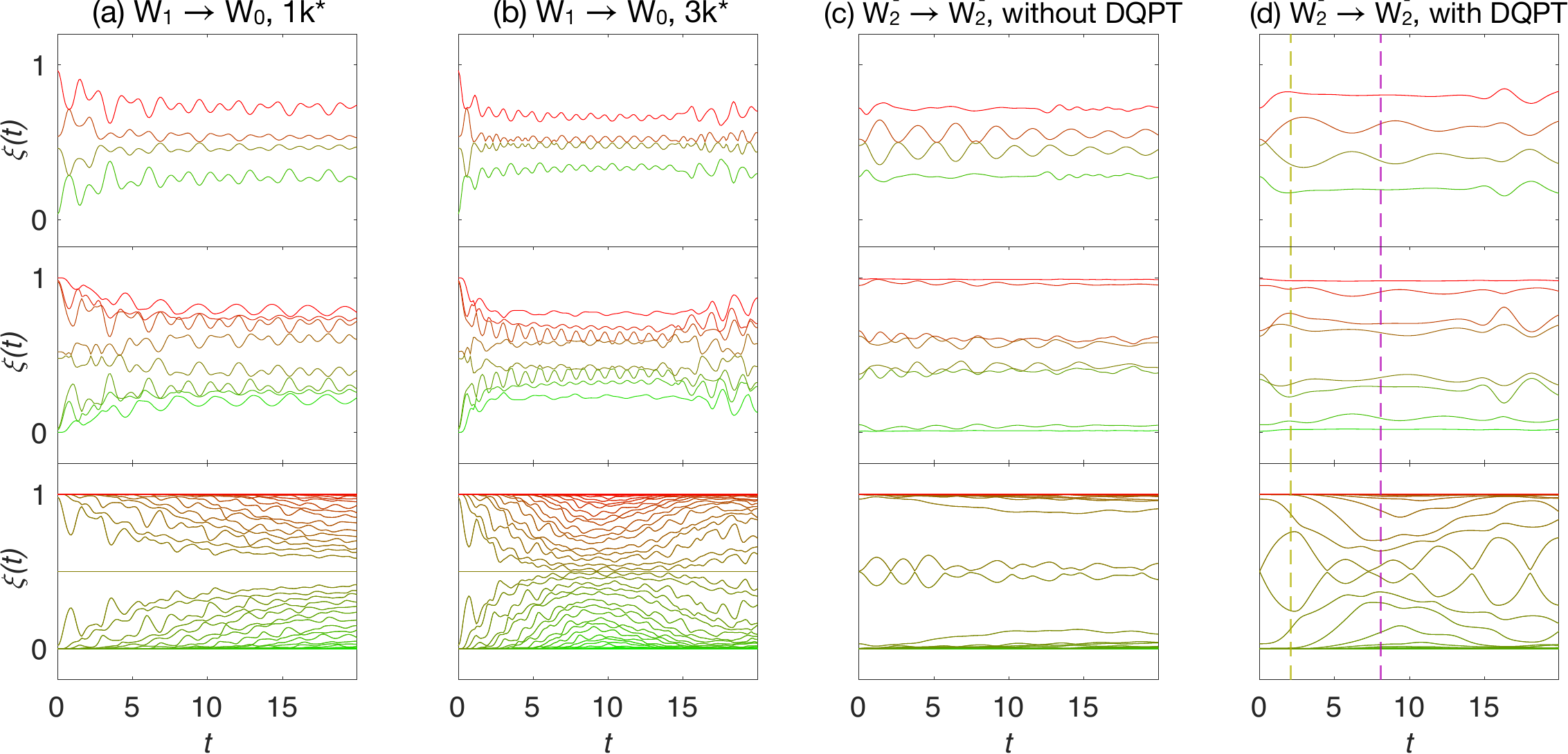}
    \caption{Time evolution of the correlation matrix spectra for quenches (a) $(-0.5,1) \rightarrow (1.01,-0.61)$, $W_1 \rightarrow W_0$ inside $1k^*$ region; (b) $(-0.5,1) \rightarrow (1.58,-1.18)$, $W_1 \rightarrow W_0$ inside $3k^*$ region; (c) $(-1.2,-1.8) \rightarrow (-0.5,-1.6)$, $W_2^- \rightarrow W_2^-$ without DQPT and (d) $(-1.2,-1.8) \rightarrow (-0.5,-0.9)$, $W_2^- \rightarrow W_2^-$ with DQPT for an $L = 80$ system. Top, middle and bottom rows represent subsystems $L_A = 2,4,40$ respectively. Colored dashed lines in (d) indicate first appearance of critical times. The first occurrence of the critical times for (a) and (b) are $t_1^*=1.1642$ and $(t_1^*,t_2^*,t_3^*)=(0.5333,0.5661,0.8962)$ respectively and are not shown in the plot.}
    \label{fig:CMS_W1W0_W2W2}
\end{figure*}


In this section, we further label the upper/lower $W_2$ phases as $W_2^{+/-}$ phase and the left/right $W_{-1}$ phases as $W_{-1}^{-/+}$ phase for clearer distinction in analysis (if not specified, we consider both). Following the analysis on EE, we study the evolution of the correlation matrix spectrum mentioned in the previous section under different quench scenarios.

In the original SSH model, it has been found that the eigen-levels of the correlation matrix cross at the Fermi level ($\xi_j=0.5$) if one quenches from the trivial $W_0$ phase to the topological $W_1$ phase and no crossings occur for quenches within the trivial phase \cite{Gong2018}. Similar phenomenon happens for the trivial to topological phase quench in a model of the same Altland-Zirnbauer BDI class as the SSH model \cite{Poyhonen2021}.  Moreover, persistent double degeneracy at the Fermi level is observed for quenches from topological to trivial phase \cite{Poyhonen2021}. The literature further proposed a dynamical quantity, termed entanglement echo, to indicate the crossings via a sudden jump in its rate function at the corresponding times. We found that the entanglement echo can behave very differently for different subsystem sizes and it does not always resemble the evolution of the Loschmidt rate in large subsystems, unlike the cases shown in Ref. \cite{Poyhonen2021} (see Appendix \ref{sec:app_C}).

The above phenomena in the correlation matrix spectrum are not limited to the mentioned quench scenarios in the extended SSH model. Figure \ref{fig:CMS_phases} shows the correlation matrix spectrum for quenches from each phase in the extended model. In addition to the consistent features found in the literature \cite{Gong2018,Poyhonen2021} for $W_1\leftrightarrow W_{0}$ phase quenches (not presented here), we also observe similar features in $W_{-1}\leftrightarrow W_0$ phase quenches, where there are level crossings in the quench from the trivial to the topological phase (Fig. \ref{fig:CMS_phases}(a)), and a persistent double degeneracy at \red{$\xi_j(t) = 0.5$}, i.e. the Fermi level for the opposite quench (Fig. \ref{fig:CMS_phases}(d)). \green{The latter can be deduced by noting that the Fermi level is doubly degenerated initially in the $W_{-1}$ phase and its evolution is a constant over time. Such a persistent degeneracy can also be observed for quenches between topological phases as shown in Fig. \ref{fig:CMS_phases}(b).} Furthermore, we notice that level crossings occur when the system is quenched from the topological $W_2$ phase to the trivial phase, as shown in Fig. \ref{fig:CMS_phases}(c). \green{Note that the Fermi level is four-fold degenerate initially in the $W_2$ phase and evolve into a pair of doubly degenerated levels until they cross at the Fermi level again. The crossing results in an instantaneous four-fold degeneracy of the Fermi level, retaining the number of edge modes in $W_2$ phases at the crossing times.}

We further investigate the correlation matrix spectrum's evolution of different subsystem sizes for quenches starting from the $W_1$ and $W_2$ phases regarding the anomalous regions found in Fig. \ref{fig:Dphasediagram}. Figures \ref{fig:CMS_W1W0_W2W2}(a) and \ref{fig:CMS_W1W0_W2W2}(b) present the correlation matrix spectrum for $W_1\rightarrow W_0$ quench in the regular and anomalous region respectively. Level crossing in the half-block's correlation matrix spectrum is observed in $3k^*$ phase but not in $1k^*$ phase (bottom panels of Figs. \ref{fig:CMS_W1W0_W2W2}(a) and \ref{fig:CMS_W1W0_W2W2}(b)). The cases for $L_A = 2$ behave the same but in a more drastic manner -- level crossings barely occur when quenched to $1k^*$ region (top panel of Fig. \ref{fig:CMS_W1W0_W2W2}(a)), whereas the near-Fermi levels frequently cross each other for quenches to $3k^*$ region, as shown in the top panel of Fig. \ref{fig:CMS_W1W0_W2W2}(b). Subsystems $L_A = 4$ seem to be insensitive to this kind of dynamical change, where there is no significant distinct feature in their correlation matrix spectra, as observed from the middle panels of Figs. \ref{fig:CMS_W1W0_W2W2}(a) and \ref{fig:CMS_W1W0_W2W2}(b). Overall, the quench to the anomalous regions triggers more frequent correlation redistributions for both very small and very large subsystems.

\begin{figure}
    \centering
    \includegraphics[width=8cm]{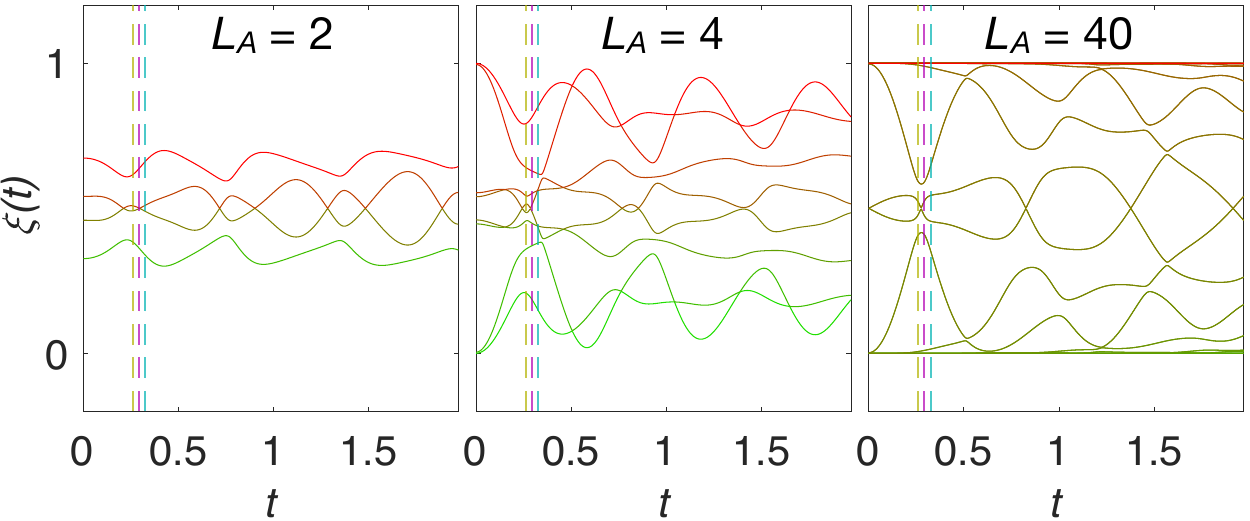}
    \caption{Time evolution of correlation matrix spectrum for the quench $(-0.5,-2) \rightarrow (-5.75,-0.5)$ for an $L = 80$ system. Colored dashed lines indicate the first occurrence of the critical times.}
    \label{fig:CMS_W2ntoWn1n_deep}
\end{figure}

The features discussed above can also be spotted in quenches within the $W_2$ phase with anomalous DQPTs. Eigen-levels in the half-block's correlation matrix cross at the Fermi level, as shown in the bottom panel of Fig. \ref{fig:CMS_W1W0_W2W2}(d). In addition, more eigen-levels away from the Fermi level approach the Fermi level as compared to its regular counterpart (bottom panel of Fig. \ref{fig:CMS_W1W0_W2W2}(c)). This suggests the quench dynamics inside the anomalous regions involves more eigen-levels, leading to a more dramatic phenomenon with the addition of the two extra critical momenta. 
On the other hand, there are a couple of differences between the $W_2^-\rightarrow W_2^-$ and $W_1\rightarrow W_0$ quenches. In particular, level crossings in half-block subsystem for $W_2^-\rightarrow W_2^-$ occur even without the emergence of DQPTs (bottom panel of Fig. \ref{fig:CMS_W1W0_W2W2}(c)). 
In Figs. \ref{fig:CMS_W1W0_W2W2}(c) and \ref{fig:CMS_W1W0_W2W2}(d), we also found a relatively steady evolution of the eigen-levels away from the fermi level for all subsystems shown. For the quench within $W_2$ phase, the evolutions are less entangled and crossings rarely occur among the eigen-levels away from the two around the Fermi level, unlike the situation in $W_1 \rightarrow W_0$ 3$k^*$. Note that these features also apply to quenches between 2 distinct $W_2$ phases ($W_2^- \leftrightarrow W_2^+$).

In the previous section, we argued that the $W_2$ and $W_{-1}$ phases have higher non-locality where the quench dynamics, especially in the vicinity of DQPTs, has their own characteristics compared to that for quenches from the topological phase $W_1$ (See Table \ref{tab:ent}). 
In Fig. \ref{fig:CMS_phases}(d), we found that the two eigen-levels (each doubly degenerated) around the Fermi level are locally the furthest away from the Fermi level near the first critical time. These levels have the highest contribution to the time evolution of the entanglement entropy since most of the other levels evolve relatively steadily. As a result, the entanglement entropy exhibits a local minimum, implying a temporary gain of information in the subsystem around the first critical time, as presented for the case of the anomalous DQPT involving class 2 $\rightarrow$ class 2 quench in the previous section.

On the contrary, one witnesses the rapid reorganization of correlation in a regular DQPT involving the class 2 $\rightarrow$ class 2 deep quench. Figure \ref{fig:CMS_W2ntoWn1n_deep} shows the time evolution of the correlation matrix for a deep quench from $W_2^-$ to $W_{-1}^-$ phase. Note that for such kind of deep quenches, the originally three distinct critical times become close to each other due to the increasingly large energy each \textit{k} mode possesses for $t_c^f \gg 1$ or $t_d^f \gg 1$. One observes level crossings around the critical times for all the subsystems concerned. Moreover, especially for $L_A=2, 4$, the upper and lower levels also have their closest approach to the Fermi level around the first critical time. Same phenomenon can be observed for quenches to $W_{-1}^+$ phase. 
This is consistent with the observation that the entanglement entropy attains a local maximum around the first critical times as shown in Fig. \ref{fig:ent-2and-1}(a). In the vicinity of DQPT, the system experiences a temporary redistribution of correlation in the subsystem, where a short instant of momentum uncertainty is introduced by the formation of quasiparticle modes at the Fermi level resulting in a restoration of the initial topological feature. This uncertainty in turn causes the loss of information to correctly describe the whole system and thus an increase in the entanglement entropy. 
\green{The features regarding the correlation matrix spectrum presented in this section can also be observed in the extended SSH model with further neighbor hoppings, as shown in Appendix \ref{sec:app_A2}.}




\section{Conclusion}
\label{sec:conclusion}
We \red{attempted to integrate several types of DQPTs into the same framework, namely, the type-B anomalous DQPTs, by comparing it with regular DQPTs } in the extended SSH model \red{with two chiral-symmetry-preserving next-nearest-neighbor hopping terms added} to the original SSH model. The introduction of these hopping terms induces two additional phases on top of $W_0$ and $W_1$, namely $W_2$ and $W_{-1}$, in equilibrium. We investigated the dynamical phase diagram by counting the number of critical momenta $n_{k^*}$ of the corresponding quench, and \red{defined the type-B anomalous DQPTs as} $n_{k^*}$ bigger than the absolute winding number difference between the pre- and postquenched phases. \red{Under this formulation, type-B anomalous DQPTs} \green{incorporate } the possibility of DQPTs occuring for quenches within the same phase \red{(both with and without crossing equilibrium phase boundary), and quenches that involve multiple $n_{k^*}$ and thus critical time scales.} \red{On top of that, we elucidated} the extra critical momenta always come in a pair due to the fact that they are complex conjugate pair of each other. The trajectories of the dynamical vector $\vec{r}_k$ helps to distinguish the nature of the regular critical momenta and the extra critical momenta, where the latter pair stays in the same loop of trajectory instead of two separate loops.

Furthermore, we classified the quench scenarios by the equilibrium entanglement entropy phase diagram, where $W_2$ and $W_{-1}$ are the more non-local (overall higher EE) phases forming class 2, while $W_0$ and $W_1$ are the less non-local (lower EE) phases forming class 1. As such, there are six scenarios of entropy dynamics quench, namely, from class 1 to class 2, and vice versa, from class 2 to itself with/without changing winding number, and from class 1 to itself with $1k^*$/ $3k^*$. We discussed the time evolution of the EE near the first DQPT regarding these six different situations. On one hand, if we quench deep inside the phase, DQPT occurs near the local maximum of EE for class 1 to class 2 and from class 2 to itself with $n_{k^*}=|\Delta W|$,  \red{regardless of the size of the }subsystem. In these quench cases, the deeper the quench to the phase, the more alignment between the time EE is maximized and the first critical time. The cause of this maximum comes from the level crossings occurred among the correlation matrix spectrum, giving the highest momentum uncertainty during the reorganization of correlation during DQPTs.

On the other hand, with an appropriate subsystem choice, the DQPT occurs near the local minimum of the entanglement entropy for quenches from class 2 to class 1 and class 2 to itself with $n_{k^*}>|\Delta W|$. The \red{variation in time of the correlation matrix spectrum} in this case is the least and the effective eigen-levels are farthest away from the Fermi level near the critical time. In addition, one might be able to realize a quench that attains lower entropy with a small subsystem near DQPT compared to the pre-quenched phase. A possible application is that one can utilize such property to perform measurement on just a small subsystem, yet still capable of capturing significant amount of information of the whole system comparing to its pre-quenched phase. For the quenches within class-1 phases with $3k^*$, it is consistent with the original SSH model, where $W_1\rightarrow W_0$ and $W_0\rightarrow W_1$ attain minimum and maximum for subsystem $L_A=1$ in the vicinity of DQPT respectively. Our results show DQPT occurs around the second and the first minimum of the EE time evolution in the $1k^*$ and $3k^*$ regions, respectively. We interpret the first minimum in EE in the $1k^*$ region as a preparation for the DQPT that emerges when entering the $3k^*$ region.


Our work may serve as an attempt to extend the study of DQPTs in topological models with more than one topological phases with distinct topological characters. The question of any physical quantities like the order parameter, whether equilibrium or dynamical, can detect the anomalous regions studied in this work remains to be explored. Topological models that happen to display \red{type-B} anomalous dynamical regions in the dynamical phase diagram are of particular interest in the future research. \red{A natural extension would be to generalize the definition of type-B anomalous DQPTs to other models of different topological invariants such as Chern number \cite{NNNSSHDQPT,Li2014} and models in higher dimensions.} \green{For example, for models without chiral symmetry where the winding number is not a good topological invariant \cite{multibandSSHDQPT}.} \red{It is also intriguing to explore the applicability of our results to generalization of unit cells instead of hopping range, which also exhibits type-B anomalous DQPTs \cite{NNNSSHDQPT}. Besides, investigating the type-B anomalous DQPTs under the influence of non-Hermitian couplings in the SSH model \cite{Nehra2024} also presents an interesting research direction.
On the other hand, the Loschmidt amplitude spectrum \cite{Wong2022,Niu2023} may also serve as a probe to study the physical effect of the extra critical momenta in topological models in the future work.}

\begin{acknowledgments}
We acknowledge financial support from Research Grants Council
of Hong Kong (Grant No. CityU 21304020), FCT through Grant No. UID/CTM/04540/2019, and City University of Hong Kong (Grant No. 9610438, No. 7006018, and No. 9680320).
\end{acknowledgments}

\appendix

\red{
\section{Type-B anomalous DQPTs in the SSH model with the next-next-nearest-neighbor hopping}
\label{sec:app_A}}

\red{
In order to check the generality of our findings presented in the main text, we here further extend the SSH model to include the next-next-nearest-neighbor hoppings. The Hamiltonian is as follow:
\begin{align}
    H &= \sum_{j = 1}^L ( t_ac_{j,A}^\dagger c_{j,B} + t_bc_{j,B}^\dagger c_{j + 1,A} \nonumber \\
    & \qquad + t_cc_{j,A}^\dagger c_{j + 1,B} + t_dc_{j,B}^\dagger c_{j + 2,A} \nonumber \\
    & \qquad +t_ec_{j,A}^\dagger c_{j + 2,B} + t_fc_{j,B}^\dagger c_{j + 3,A} +  \text{h.c.} ),
\end{align}
where $t_e$ and $t_f$ characterize the corresponding next-next-nearest-neighbor hopping strength.} \green{Note that the Hamiltonian still preserves the chiral symmetry.} \red{The further neighbor hopping opens up the possibilities for the system favoring topological phases with winding numbers $W=3$ and $W=-2$ (we label the phase as $W_3$ and $W_{-2}$ respectively). These two new phases neither belongs to class 2 nor class 1 but a class with higher degree of non-locality according to the properties of the equilibrium EE (not shown here). We label this class as class 3 in the following.}

\red{The Bloch vector takes the following form:
\begin{equation}
    \begin{aligned}
        d_x(k) &= t_a + ( t_b + t_c )\cos k + (t_d+t_e)\cos( 2k ) +t_f \cos(3k),\\
        d_y(k) &= ( t_b - t_c )\sin k + (t_d-t_e)\sin( 2k )+t_f\sin (3k), \\
        d_z(k) &= 0.
    \end{aligned}
\end{equation}
This is equivalent to the generalized two-band model with long-range hopping truncated to $m = 3$ in Ref. \cite{accident2015}. The orthogonality condition $\vec{d}_i(k^*) \cdot \vec{d}_f(k^*)=0$ gives a quintic equation in the cosines and there are at most five critical momenta in this model.}


\begin{figure}
    \centering
    \includegraphics[width=8.5cm]{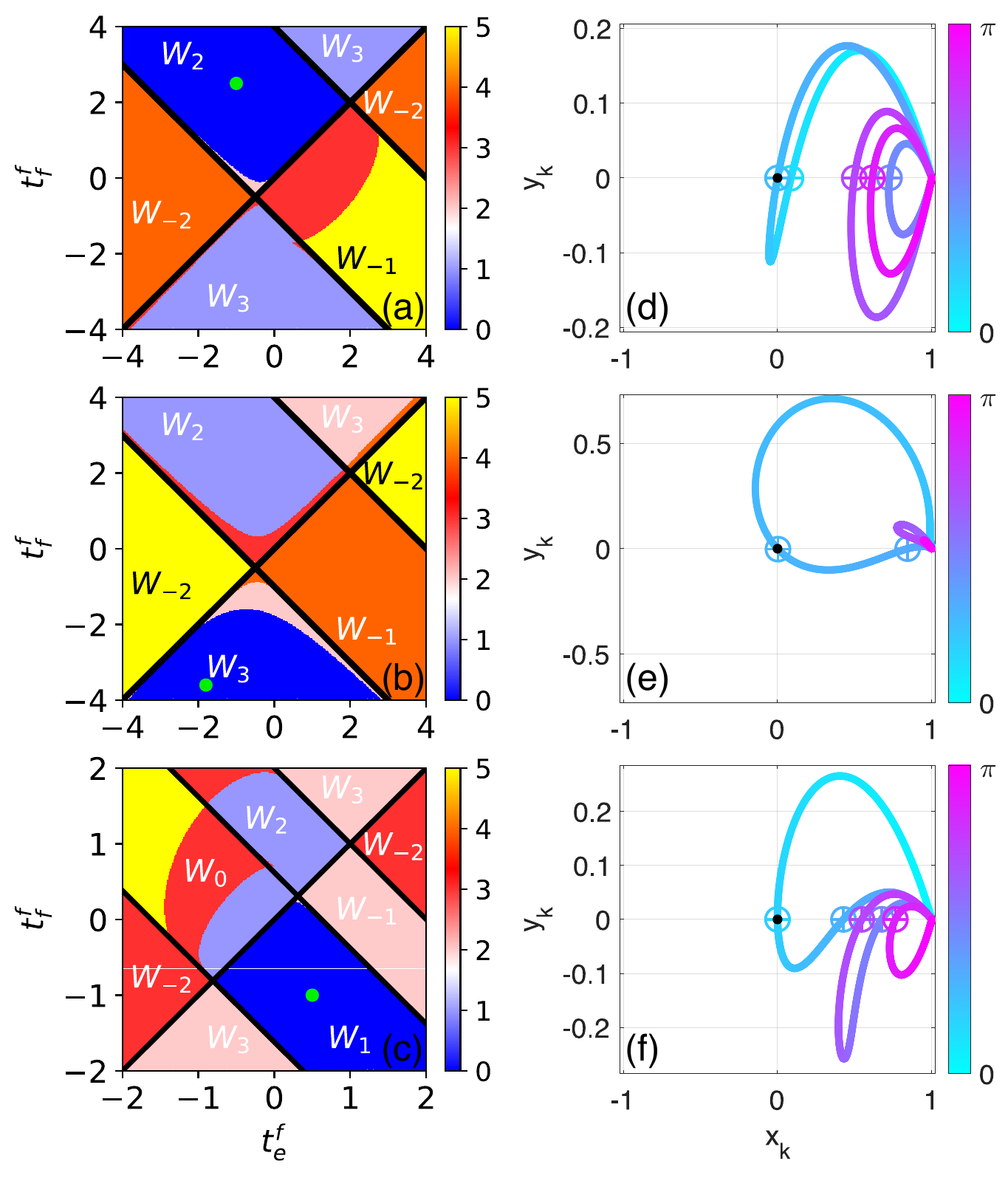}
    \caption{\red{Left panels: The dynamical phase diagrams for initial state (green dot) in phase (a) $W_2$ $(t_e^i = -1,t_f^i = 2.5)$ and (b) $W_3^-$ $(t_e^i = -1.8,t_f^i = -3.6)$ with $t_a = t_b = t_c = t_d = -1$,; (c) $W_1$ $(t_e^i = 0.5,t_f^i = -1)$ with $t_a = t_b = -1,t_c = t_d = 0$ for the SSH model with next-next-nearest-neighbor hopping. Color bars show the number of critical momenta. The equilibrium phase boundaries are indicated by black solid lines, and the phases are denoted by $W_n$, where $n\in \{-2,-1,0,1,2,3\}$. Right panels: Trajectories of $\vec{r}_k$ at the earliest critical time of quenches (d) $W_2 \rightarrow W_{-1}$ $(t_e^i = -1,t_f^i = 2.5) \rightarrow (t_e^f = 2.25,t_f^f = -1.05)$ inside $5k^*$ region, (e) $W_3^- \rightarrow W_3^-$ $(t_e^i = -1.8,t_f^i = -3.6) \rightarrow (t_e^f = -0.5,t_f^f = -1.4)$ with DQPT and (f) $W_1 \rightarrow W_0$ $(t_e^i = 0.5,t_f^i = -1) \rightarrow (t_e^f = -1.6,t_f^f = 1.1)$ inside $5k^*$ region. Black dots indicate the origin. The ``$\oplus$" signs pin the corresponding critical momenta. Here $L = 8000$.}}
    \label{fig:rk_tetf}
\end{figure}

\red{The dynamical phase diagram for $t_a=t_b=t_c=t_d=-1$ is shown in Figs. \ref{fig:rk_tetf}(a) and \ref{fig:rk_tetf}(b) with prequenched phase as $W_2$ and the lower $W_3$ phase, respectively.} 
\green{Similar to the dynamical phase diagram in Figs. \ref{fig:Dphasediagram}(b) and \ref{fig:Dphasediagram}(c) for $t_e = t_f = 0$, anomalous regions with extra pair(s) of $k^*$'s are found for quench $W_2\rightarrow W_{-1}$ and $W_3\rightarrow W_3$.} \red{We notice a couple of differences regarding the initial phase as the lower $W_3$ phase under the presence of the longer-range hoppings (compare Fig. \ref{fig:rk_tetf}(b) and Fig. \ref{fig:Dphasediagram}(c)): There is a small anomalous region near the phase boundaries $t_f = -t_e - 1$ and $t_f = t_e$ of $W_2$, and also $t_f = t_e$ of the upper $W_3$; a tiny anomalous region of $4k^*$'s besides the $2k^*$ region appears inside the lower $W_3$ phase near the multicritical point $(t_e^c,t_f^c) = (-0.5,-0.5)$. Nevertheless, the critical momenta pairs in $\vec{r}_k$} \green{of the anomalous DQPTs} \red{behave the same as described in the main text, i.e. they stay in one loop as shown in Figs. \ref{fig:rk_tetf}(d) and \ref{fig:rk_tetf}(e). Additionally, the dynamical phase diagram for $t_a = t_b = -1,t_c = t_d = 0$ from the $W_1$ phase is also shown in Fig. \ref{fig:rk_tetf}(c).} \green{For the quench to the $W_0$ phase, there are two anomalous regions, one with $3k^*$ and one with $5k^*$, observed. The $\vec{r}_k$ trajectory for a quench to the $5k^*$ region is depicted in Fig. \ref{fig:rk_tetf}(f).} \green{It clearly shows two pairs of extra $k^*$'s, with each pair staying in one loop, again validating the complex-conjugate-pair statement presented in the main text.}

\begin{figure}
    \centering
    \includegraphics[width=8.5cm]{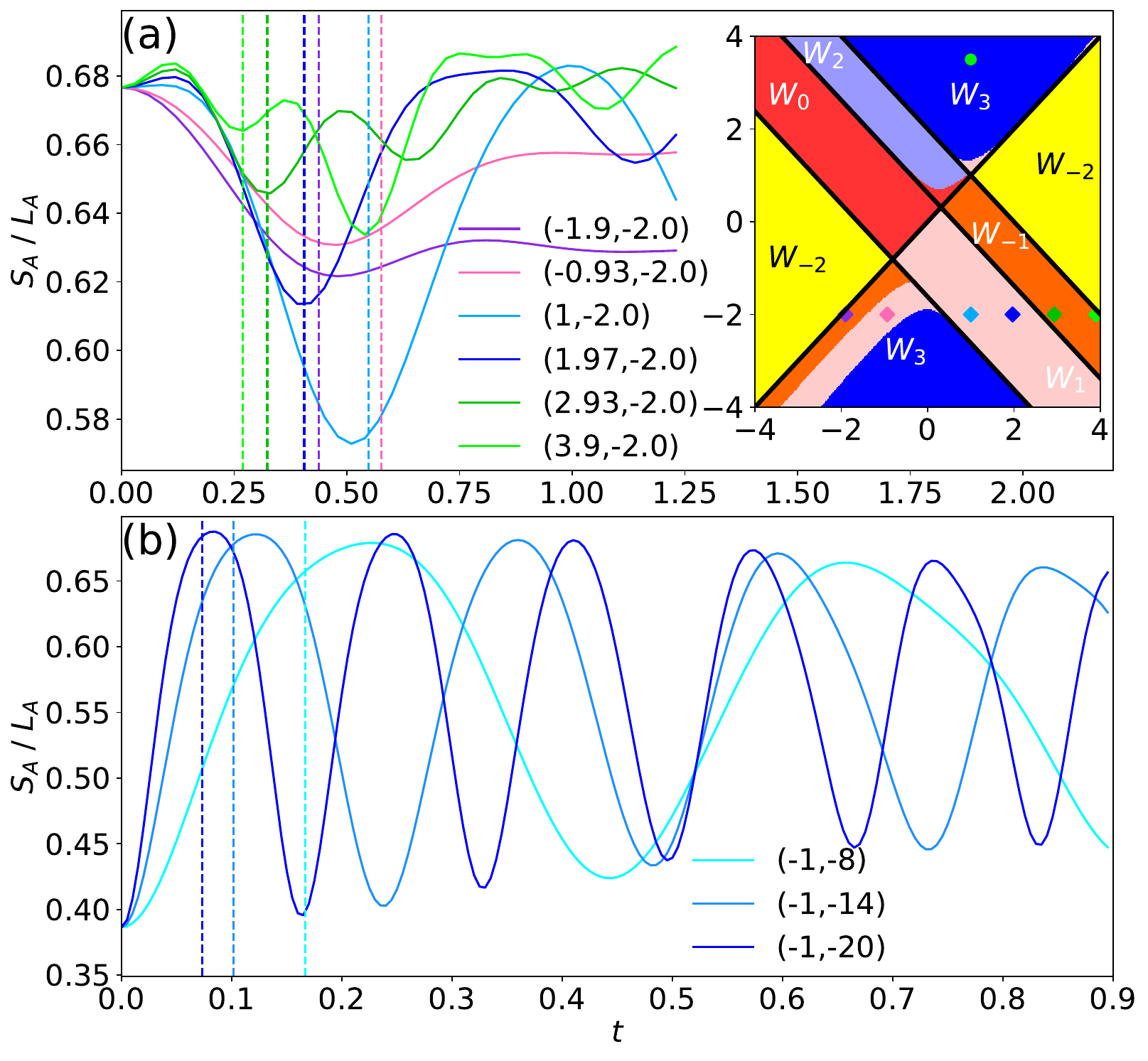}
    \caption{\red{Time evolution of the entanglement entropies per unit cell with subsystem $L_A=3$, quenching from (a) $W_3$ to some of the phases in class 1,2,3 and (b) $W_{-1}$ to the deep $W_3$ phase. The vertical dashed lines show the first critical time for each of the quench cases. Quenches in (a) and (b) are $(t_a,t_b,t_c,t_d,t_e,t_f)=(-1,-1,0,0,t_e^i=1,t_f^i=3.5)\rightarrow(-1,-1,0,0,t_e^f,t_f^f)$ and $(t_a,t_b,t_c,t_d,t_e,t_f)=(-1,-1,0,0,t_e^i=1,t_f^i=0)\rightarrow(-1,-1,0,0,t_e^f,t_f^f)$ respectively, 
    where the exact values of $t_e^f$ and $t_f^f$ are indicated in the legends. The $n_{k^*}$ dynamical phase diagram for the quench in (a) is shown in the corresponding inset. The same color scale as in Fig. \ref{fig:rk_tetf} is adopted for the dynamical phase diagram. Here $L=80$.}}
    \label{fig:ent_tetf}
\end{figure}

\red{
\subsection{Entanglement entropies}
\label{sec:app_A1}}

\red{From Table \ref{tab:ent}, note that for a phase in Class 2 quenching to that in Class 1, a local minimum of the EE can be achieved in the vicinity of DQPT with a suitable choice of the subsystem. The finding may also be generalized to other cases with quench from a higher to a lower classes, and this can be confirmed by Fig. \ref{fig:ent_tetf}(a), where the EE for quenches from $W_3$ to phases in all the classes, including $W_1$ (Class 1), $W_{-1}$ (Class 2), and $W_3$ (Class 3), are presented. It clearly shows the critical time occurs near the EE's local minima for quench from $W_3$} \green{to $W_1$ and $W_{-1}$. In the main text, we also find that the anomalous DQPT resulted from} \red{the quench from $W_2$ or $W_{-1}$ to itself exhibits a local minimum in the EE with a suitable subsystem choice (see Table \ref{tab:ent}, Figs. \ref{fig:ent-2and-1}(b) and \ref{fig:ent-2and-1}(c)). The quench for $W_3$ and $W_{-2}$ should also follow similar pattern and this is validated by the pink and purple curves in Fig. \ref{fig:ent_tetf}(a). The EE for the $W_3\rightarrow W_3$ quench exhibits a local minimum near DQPT, regardless of $n_{k^*}=2$ or $n_{k^*}=4$. We have verified the above also holds for quench initially from the $W_{-2}$ phase with a subsystem of $L_B=3$ (not shown here).}

\red{In the main text, we also find that local maxima in the EE can be attained around DQPT for deep quenches from Class 1 to Class 2 (see Fig. \ref{fig:ent-betweenClass}(b)), or from Class 2 to Class 2 with a regular DQPT (see Fig. \ref{fig:ent-2and-1}(a)). One infers that deep quenches from a lower to a higher class or between phases in the same class with regular DQPTs will also achieve a local maximum in EE around the critical time. In order words, we expect deep quenches from $W_{-1/0/1/2}$ to $W_{-2/3}$ or $W_{-2}/W_3$ to $W_3/W_{-2}$ to show local maxima in EE around DQPTs. 
Figure \ref{fig:ent_tetf}(b) shows a deep phase quench of $W_{-1}\rightarrow W_3$, where the darker the color of the curve is, the deeper the final parameter is inside the phase. We indeed observe maxima of EE around DQPTs. In addition, the deeper into the $W_3$ phase, the more the EE maxima and the critical time align. This is consistent with what we have obtained from the inset of Fig. \ref{fig:ent-2and-1}(a). }

\begin{figure*}
    \centering
    \includegraphics[width=16cm]{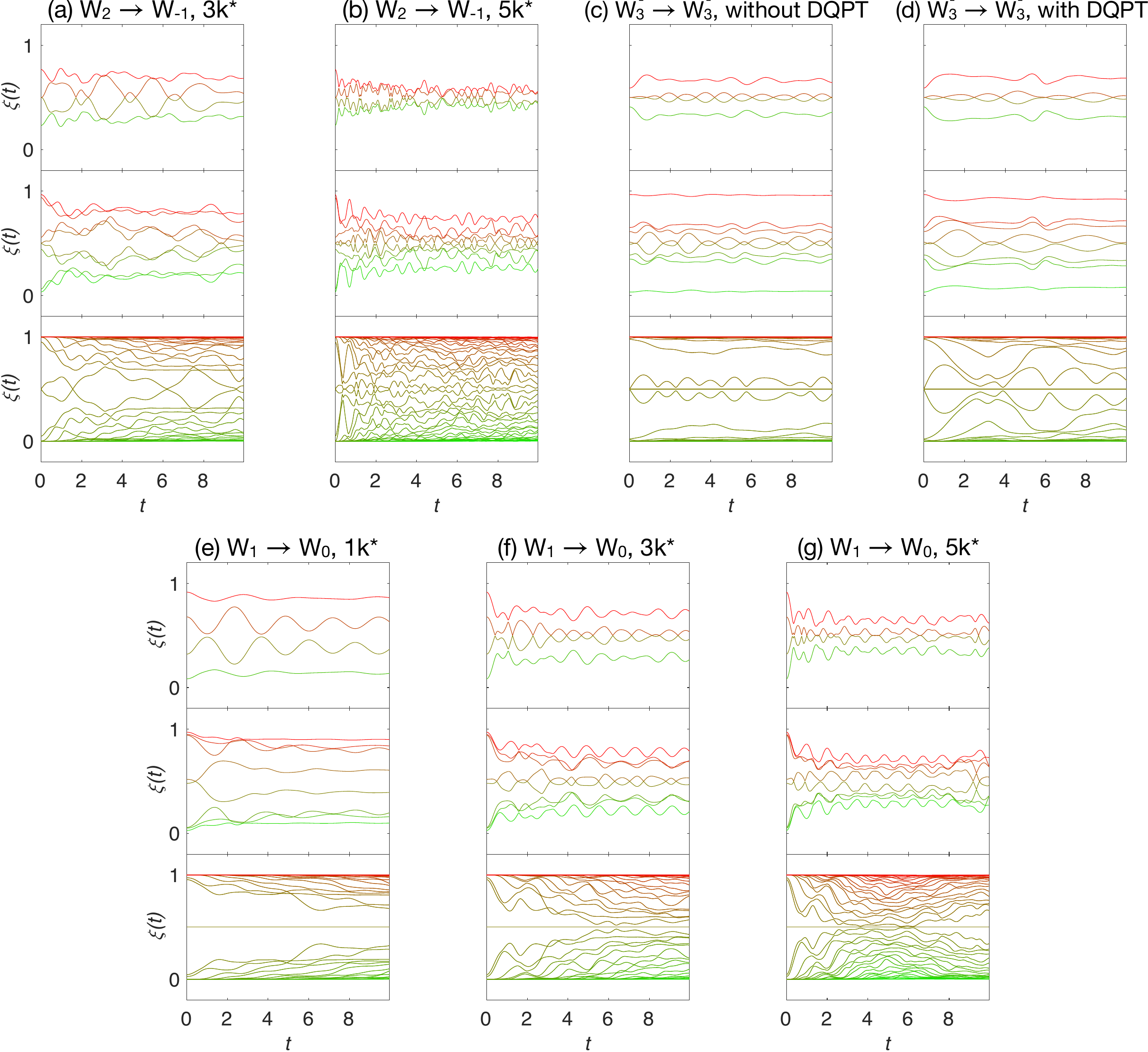}
    \caption{\red{Time evolution of the correlation matrix spectra for an $L = 80$ system with (a-d) $t_a = t_b = t_c = t_d = -1$ and (e-g) $t_a = t_b = -1,t_c = t_d = 0$  (see Fig. \ref{fig:rk_tetf}(a), \ref{fig:rk_tetf}(b) and \ref{fig:rk_tetf}(c) for the dynamical phase diagrams respectively). The quench cases are (a) $(t_e^i = -1,t_f^i = 2.5) \rightarrow (t_e^f = 1.9,t_f^f = 0.9)$, $W_2 \rightarrow W_{-1}$ inside $3k^*$ region; (b) $(t_e^i = -1,t_f^i = 2.5) \rightarrow (t_e^f = 3.9,t_f^f = -0.8)$, $W_2 \rightarrow W_{-1}$ inside $5k^*$ region; (c) $(t_e^i = -1.8,t_f^i = -3.6) \rightarrow (t_e^f = -0.5,t_f^f = -2.1)$, $W_3^- \rightarrow W_3^-$ without DQPT, (d) $(t_e^i = -1.8,t_f^i = -3.6) \rightarrow (t_e^f = -0.5,t_f^f = -1.4)$, $W_3^- \rightarrow W_3^-$ with DQPT; (e) $(t_e^i = 0.5,t_f^i = -1) \rightarrow (t_e^f = -0.5,t_f^f = -0.1)$, $W_{1} \rightarrow W_0$ inside $1k^*$ region; (f) $(t_e^i = 0.5,t_f^i = -1) \rightarrow (t_e^f = -1.05,t_f^f = 0.5)$, $W_{1} \rightarrow W_0$ inside $3k^*$ region; (g) $(t_e^i = 0.5,t_f^i = -1) \rightarrow (t_e^f = -1.6,t_f^f = 1.1)$, $W_{1} \rightarrow W_0$ inside $5k^*$ region. Top, middle and bottom rows represent subsystems $L_A = 2,4,40$ respectively.}}
    \label{fig:CMS_tetf}
\end{figure*}

\red{
\subsection{Correlation matrix spectra}
\label{sec:app_A2}}

\red{We adopt the notation for the upper/lower $W_3$ phases as $W_3^{+/-}$ in this section.  Figure \ref{fig:CMS_tetf} compares the time evolution of the correlation matrix spectrum between quenches to type-B anomalous and regular regions. The initial quench parameters in Figs. \ref{fig:CMS_tetf}(a)-\ref{fig:CMS_tetf}(b), \ref{fig:CMS_tetf}(c)-\ref{fig:CMS_tetf}(d), and \ref{fig:CMS_tetf}(e)-\ref{fig:CMS_tetf}(g) are the same as that in the dynamical phase diagram shown in Figs. \ref{fig:rk_tetf}(a), \ref{fig:rk_tetf}(b), and \ref{fig:rk_tetf}(c), respectively. The eigen-levels for $W_2 \rightarrow W_{-1}$ with anomalous DQPT (Fig. \ref{fig:CMS_tetf}(b)) is more entangled and concentrated around the Fermi level, and at the same time the quench triggers slightly more level crossings as compared to the regular case (Fig. \ref{fig:CMS_tetf}(a)). This is similar to the observations for the $W_1 \rightarrow W_0$ case (see Figs. \ref{fig:CMS_W1W0_W2W2}(a) and \ref{fig:CMS_W1W0_W2W2}(b)) in the main text. Note however that for the half-block's correlation matrix spectrum, level crossing at the Fermi level are found in both the regular and the type-B anomalous cases (the bottom panel of Figs. \ref{fig:CMS_tetf}(a) and \ref{fig:CMS_tetf}(b)), unlike the case for $W_1 \rightarrow W_0$ where quenches to the regular region (Fig. \ref{fig:CMS_W1W0_W2W2}(a)) do not trigger crossings for a reasonably long time.} \green{The difference may arise from the fact that quench between two topological phases is involved for $W_2 \rightarrow W_{-1}$, as compared to the quench between a topological phase and a trivial phase for $W_1 \rightarrow W_0.$  On the other hand, if one considers the quench from a topological $W_1$ phase to a trivial $W_0$ phase in the further extended SSH model, as shown in Figs. \ref{fig:CMS_tetf}(e)-\ref{fig:CMS_tetf}(g), the distinctive features between the anomalous and the regular DQPTs are found to be consistent with those presented in Sec. \ref{sec:CMS}. Namely, persistent double degeneracy at the Fermi level with the degenerate levels gapped from the rest in the regular case ($1k^*$), and the levels tend to mix and cross for both the $3k^*$ and $5k^*$ anomalous cases.} 

\red{The quenches within the $W_3^-$ phases shown in Figs. \ref{fig:CMS_tetf}(c)-\ref{fig:CMS_tetf}(d) also resemble the dynamics of the correlation matrix spectrum for quenches within the $W_2^-$ phase shown in Figs. \ref{fig:CMS_W1W0_W2W2}(c)-\ref{fig:CMS_W1W0_W2W2}(d).} \green{For the subsystem of $L_A = 2,4$, the spectrum has more crossings in the regular case as compared to the anomalous case. Furthermore, the eigen-levels around the Fermi level for the half-block subsystem are gapped from the rest in the regular case, whereas the eigen-levels away from the Fermi level are more involved in the anomalous case. We also note that there is a persistent double degeneracy at the Fermi level in the half-block correlation matrix spectrum for the quench from $W_3$, which is absent in the case of quench from $W_2$ in the main text. The subsystem recovers the six-fold degeneracy temporarily at the time where level crossing at the Fermi level occurs.} \red{The persistent double degeneracy in the half-block subsystems is in fact guaranteed for quenches from a phase with odd winding number. Note that the eigen-levels whose eigenvalues are neither 0 nor 1 are doubly degenerate. Since initially the phase has $2W$ edge states, the phases with odd winding number must have a pair of eigen-levels left which does not belong to either upper or lower branch of the spectrum, given the spectrum are symmetric along the Fermi level due to the presence of particle-hole symmetry. That one pair has to stay at the Fermi level, leading to a persistent double degeneracy happened in quenches from odd-winding-number phases.}

\section{Type-A anomalous DQPTs in the extended SSH model}
\label{sec:app_B}

\begin{figure}[b]
    \centering
    \includegraphics[width=8.5cm]{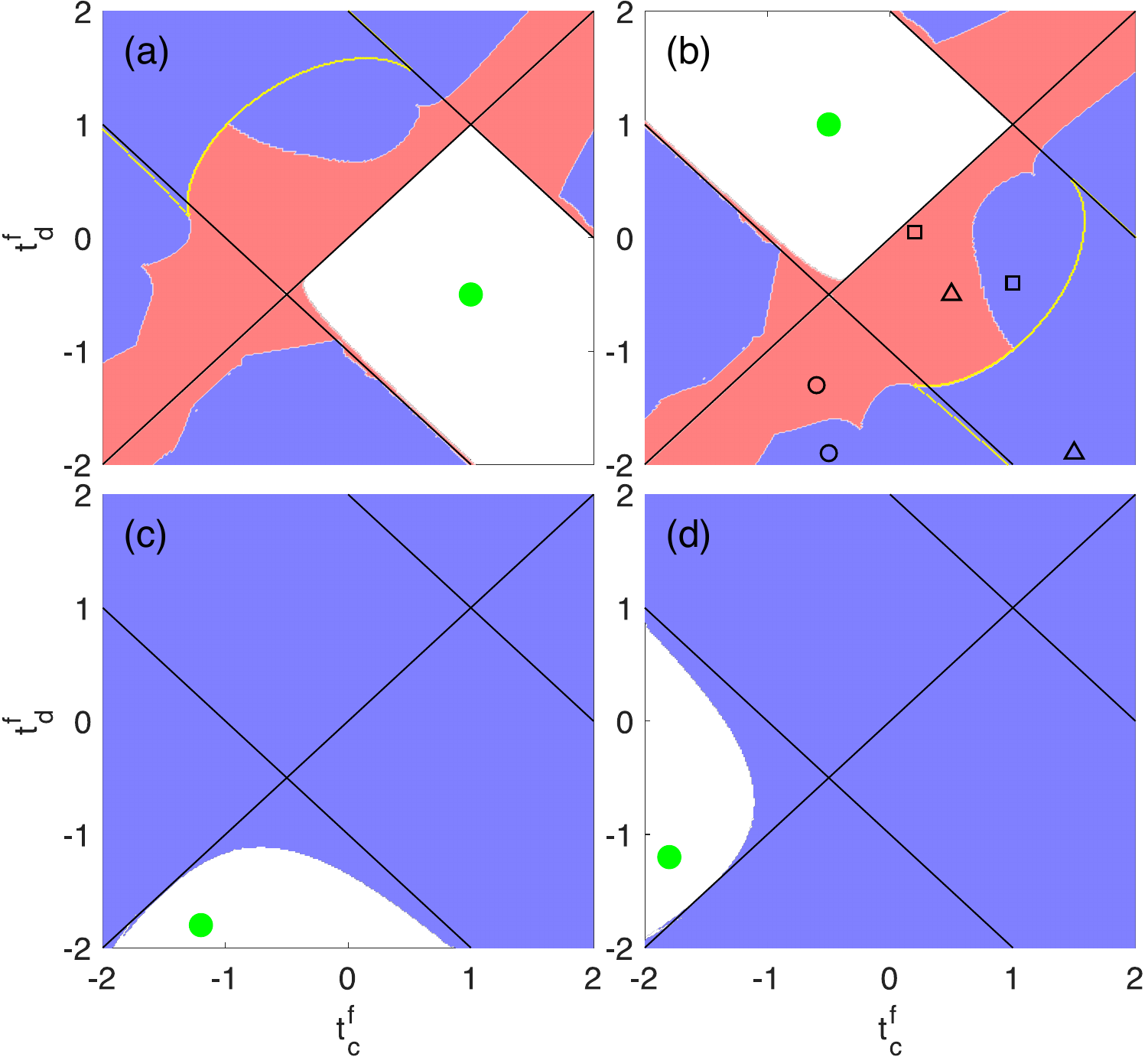}
    \caption{Dynamical phase diagram of type-A anomalous DQPTs for the same initial phases as in Fig. \ref{fig:Dphasediagram} in an $L = 10000$ system. Red regions represent type-A anomalous DQPTs, blue regions represent regular DQPTs and white regions exhibit no DQPTs. The yellow boundaries in (a) and (b) enclose the type-B anomalous regions for the corresponding phases. Green dots indicate the initial points for each diagram. The black hollow symbols in (b) pin the respective final points of the quenches shown in Fig. \ref{fig:ADQPT_LR_W1}.}
    \label{fig:ADQPT_diag}
\end{figure}

Type-A anomalous DQPTs were first discovered back in 2017 on the study of quench and dynamical phase transitions in the transverse-field Ising model with long-range interactions, where a minimum in the Loschmidt rate function occurs sooner than the first critical time (the first occurrence of DQPT) \blue{\cite{Corps2022,Halimeh2017,Homrighausen2017,Zauner-Stauber2017,Halimeh2021,Corps2023,Nicola2021,Halimeh2020,Hashizume2022,Osborne2024}}. In the extended SSH model, which is a topological model, we observe similar phenomenon for quenches between two different phases. Figure \ref{fig:ADQPT_diag} shows the dynamical phase diagram for the type-A anomalous DQPTs in different initial phases, with the red regions indicating the regions where type-A anomalous DQPTs occur. First notice that there is no type-A anomalous DQPT happening for quenches from both $W_2$ and $W_{-1}$ phases, as seen from Figs. \ref{fig:ADQPT_diag}(c) and \ref{fig:ADQPT_diag}(d) where the red region is absent. In other words, type-A anomalous DQPTs only occur for quenches from class-1 ($W_1,W_0$) phases (Figs. \ref{fig:ADQPT_diag}(a) and \ref{fig:ADQPT_diag}(b)).

\begin{figure}
    \centering
    \includegraphics[width=7cm]{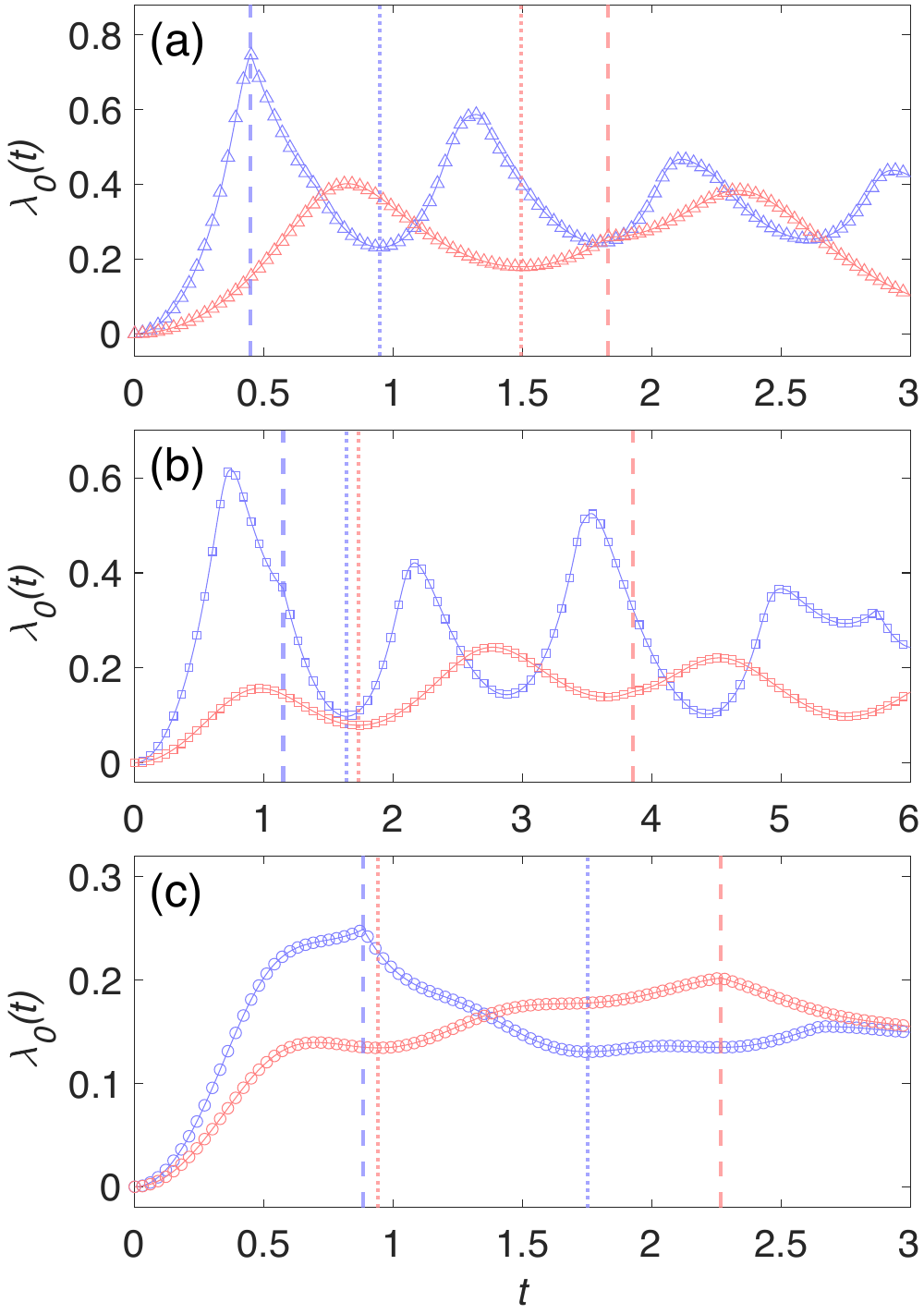}
    \caption{Rate functions of (a) $t_c^f = 0.5,t_d^f = -0.5$ (red line) and $t_c^f = 1.5,t_d^f = -1.9$ (blue line), (b) $t_c^f = 0.2,t_d^f = 0.05$ (red line) and $t_c^f = 1,t_d^f = -0.4$ (blue line), (c) $t_c^f = -0.6,t_d^f = -1.3$ (red line) and $t_c^f = -0.5,t_d^f = -1.9$ (blue line) from $W_1$ phase $t_c^i = -0.5,t_d^i = 1$ of an $L = 10000$ system. The initial point and the final points are indicated by the green circle and the corresponding hollow symbols in Fig. \ref{fig:ADQPT_diag}(b), respectively. Red and blue lines refer to \red{type-A} anomalous and regular DQPTs respectively. Colored dashed lines mark the first occurrence of DQPTs while colored dotted lines indicate the times the first minimum of LRs occur.}
    \label{fig:ADQPT_LR_W1}
\end{figure}

 In Figs. \ref{fig:ADQPT_diag}(a) and \ref{fig:ADQPT_diag}(b), the type-B anomalous ($3k^*$) regions observed in Figs. \ref{fig:Dphasediagram}(a) and \ref{fig:Dphasediagram}(b) are outlined by the yellow lines. We find that a portion of the type-B anomalous DQPT phase boundary is shared with the type-A anomalous DQPT phase boundary.
As one crosses this dynamical phase boundary from the $1k^*$ region to the $3k^*$ region, one of the two extra $k^*$'s can result in a critical time ($t^*_B$) much earlier than the original first critical time in the $1k^*$ region ($t^*_A$). As an example, Fig. \ref{fig:ADQPT_LR_W1}(a) shows the LR for quenches from the $W_1$ phase to two final points indicated by the hollow triangles in Fig. \ref{fig:ADQPT_diag}(b). The red curve shows the case in the $1k^*$ region with type-A anomalous DQPT while the blue curve shows the case in the $3k^*$ region without type-A anomalous DQPT. The first smooth peak in the red curve might be considered as a pre-development of the nonanalytic peak at $t^*_B$, which arises from one of the two extra $k^*$'s when entering the $3k^*$ region, in the blue curve. If $t^*_A$ occurs much later than $t^*_B$, the LR may experience a minimum before reaching the nonanalyticity at $t^*_A$, and thus results in a type-A anomalous DQPT. However, the blue region where type-A anomalous DQPT is absent can also be found inside the $1k^*$ region, where the critical time is relatively small, as illustrated with the LR in Fig. \ref{fig:ADQPT_LR_W1}(b).


One notices the concentration of the red regions around the $t_d = t_c$ equilibrium phase boundary in Figs. \ref{fig:ADQPT_diag}(a) and \ref{fig:ADQPT_diag}(b). Some selected quenches in these regions are demonstrated as red lines in Figs. \ref{fig:ADQPT_LR_W1}(b) and \ref{fig:ADQPT_LR_W1}(c). For the quench cases near this boundary, the critical times increase as the final points approach the boundary. Plus the closer the final points to the boundary, the more the number of local minima are built up in the LRs, these DQPTs have to be type-A anomalous. In addition to this, the minimum achieved by these quenches in the extended SSH model is not the same as those studied in Ref. \blue{\cite{Corps2022,Halimeh2017,Homrighausen2017,Zauner-Stauber2017,Halimeh2021,Corps2023,Nicola2021,Halimeh2020,Hashizume2022,Halimeh2021,Halimeh2020,Hashizume2022,Osborne2024}}. Namely, the minima shown in the literature are in fact zeros of the LRs, while the minima presented in here have a finite value. The latter suggests the system at this time  still deviates from the initial state even when the LR achieves a local minimum. This qualitative difference may be attributed to the topological characters of the present model as compared to a non-topological spin model. Further investigation into the correlation, if any, between the two types of anomalous DQPTs, and the time evolution of the order parameters will be an interesting future work.

\section{Entanglement echo rate function}
\label{sec:app_C}

\begin{figure}
    \centering
    \includegraphics[width=8.5cm]{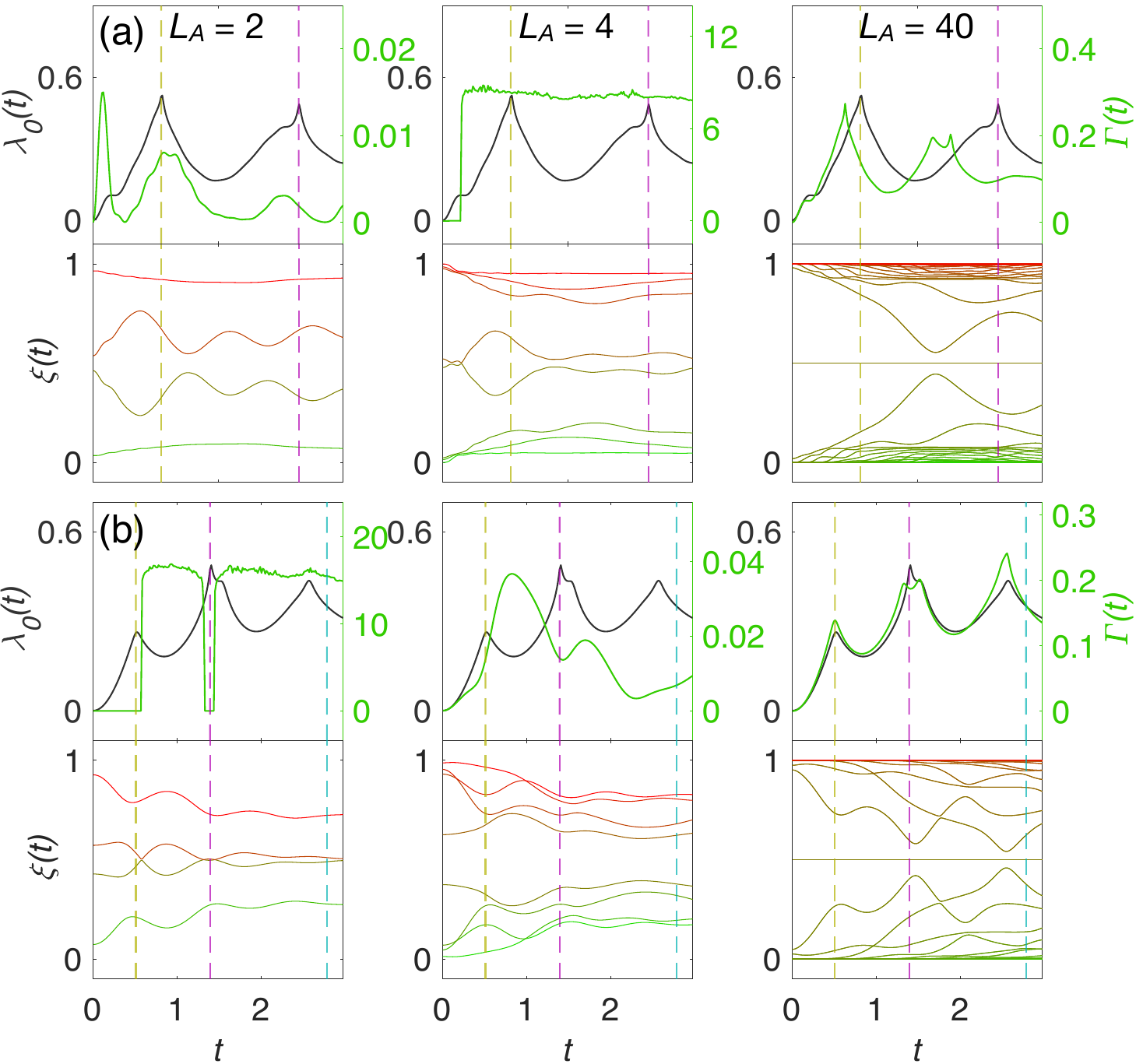}
    \caption{(Top) Loschmidt (black) and entanglement (green) rate functions, (bottom) correlation matrix spectrum of the corresponding subsystems for a quench from (a) $W_1$ to $W_{-1}^-$ $(-0.5,1) \rightarrow (-0.97,-0.63)$, and (b) $W_{-1}^-$ to $W_2^+$ $(-1.8,-1.2) \rightarrow (1,1.8)$ in an $L = 80$ system. Colored dashed lines indicate first appearance of critical times.}
    \label{fig:LRGamma_noMatch}
\end{figure}

Entanglement echo signals level crossings at the Fermi level ($\xi_j=0.5$) in the correlation matrix spectrum. It is defined as the overlap of the instantaneous entanglement ground state onto the initial entanglement ground state
\begin{equation}
    \mathcal{E}(t) = \det\braket{\xi_i(0) | \xi_j(t)}
\end{equation}
and the corresponding rate function $\Gamma(t) = -\ln[|\mathcal{E}(t)|^2] / L_\Omega$ with $L_\Omega$ being the subsystem size \cite{Poyhonen2021}. In the noninteracting picture, the Slater determinant of the eigenstates of \textit{$C_{\mu\nu}(t)$} with eigenvalues $0.5 \leq \xi_j(t) \leq 1$ gives the instantaneous entanglement ground state. $\Gamma(t)$ was shown to resemble the evolution of the Loschmidt rate for a subsystem comparable to the half-block of the full system in Ref. \cite{Poyhonen2021}. In particular, a sudden jump indicating a level crossing at the Fermi level of the correlation matrix spectrum occurs besides the usual critical time $t^*$, marking the so-called entanglement transition happening inside the subsystem. In our work, we further observe this kind of transitions for smaller subsystems as well as the half-block case. However, we realize that the $\Gamma(t)$ does not always resemble the evolution of the Loschmidt rate even with the half-block subsystem.

Figure \ref{fig:LRGamma_noMatch} serves as a demonstration of the different evolutions of $\Gamma(t)$ compared to that of Loschmidt rate. An immediate distinction can be seen in the right panel of Fig. \ref{fig:LRGamma_noMatch}(a), which shows a quench from $W_1$ to $W_{-1}^-$ phase and a half-block subsystem is considered. The two transitions deviates further when it comes to the second critical time, where the nonanalytical peak of the $\Gamma(t)$ occurs much earlier than that of the Loschmidt rate. Plus the double cusps in the second peak makes the entire evolution of $\Gamma(t)$ part with the Loschmidt rate function. This phenomenon happens occasionally when the quench involves class-2 phases, although often the entanglement echo rate function evolves similarly to that of Loschmidt rate function like the one shown in Fig. \ref{fig:LRGamma_noMatch}(b) right panel. 

The situation becomes more dramatic when the subsystems shrink further and further, for which the cases for $L_A = 2,4$ are shown in the left and the middle panels in Fig. \ref{fig:LRGamma_noMatch}. Particularly for $L_A = 2,4$ cases, large vertical jumps, either up or down, in $\Gamma(t)$ are observed right at the crossing times. From the definition of the entanglement echo, the overlap of the instantaneous entanglement ground state onto the initial entanglement ground state quickly drops to near zero once the two eigen-levels cross at the Fermi level. At any other instants $\Gamma(t)$ evolves relatively stably. The apparent difference in the magnitude of jumps in small and large subsystems can be attributed to the fact that in small subsystems, there are only a few eigenstates constituted in the entanglement ground state. A level crossing causes a relatively big change to the components of the entanglement ground state, leading to a larger difference between the initial and instantaneous entanglement ground state. On the other hand, the jumps for large subsystems turn out to be smaller, like the one stated in Ref. \cite{Poyhonen2021}, because a level crossing merely changes a tiny portion of the entanglement ground state under the enormous spectrum the subsystem has. The overall entanglement structure of the ground state still has changed, but not as dramatic as those for small subsystems.

\end{document}